\documentclass[12pt]{article}

\usepackage{graphicx}% Include figure files
\usepackage{bm}% bold math
\usepackage[utf8]{inputenc}
\usepackage[T1]{fontenc}
\usepackage{mathptmx}
\usepackage{etoolbox}
\usepackage{amsmath} %% for \text{} in equations
\usepackage{amssymb}
\usepackage{subcaption} %% to combine two or more subfigures in one figure
\usepackage{bbm} %% for imaginary unit

\begin{document}

\title{Eigenvalues of strictly regular Hall-plates}
\author{Udo Ausserlechner, https://orcid.org/0000-0002-8229-9143}
\date{\today}
\maketitle

\begin{abstract}
\noindent This work is about uniform, plane, singly connected, strictly regular Hall-plates with an arbitrary number of peripheral contacts exposed to a uniform magnetic field of arbitrary strength. The strictly regular symmetry is the highest possible degree of symmetry, and it is found in commercial Hall-plates for magnetic field sensors or circulators. It means that all contacts and contact spacings are equally large, if the Hall-plate is mapped conformally to the unit disk. The indefinite conductance matrices of such Hall-plates are circulant matrices, whose complex eigenvalues can be computed in closed form. It is shown how to express the conductance and resistance matrices of these Hall-plates, how to compute their equivalent resistor circuit, their Hall-output voltages or currents, their signal-to-thermal noise ratio, and their power as functions of the eigenvalues. It is also proven that the noise efficiency of strictly regular Hall-plates with many contacts can be up to 112\% better than for conventional Hall-plates with four contacts, and it is explained why their optimal biasing uses patterns of supply voltages or currents, which vary sinusoidally along their boundary.
\end{abstract}

\maketitle

\section{Introduction}
\label{sec:intro}
Hall-plates are used in huge quantities (up to billions) to measure magnetic fields in linear and rotational position sensors \cite{IFX-lin-Hall2006} in industrial and automotive applications (torque control of motors, control of combustion engines, anti-lock breaking systems, headlight levelling, indirect tyre pressure monitoring, steering angle sensor, end point detection of window lifters). In recent years they have been used increasingly to measure electric current with galvanic isolation via the magnetic field accompaning the current \cite{TUdelft-2023,IFX-cur-2012}. The strong points of modern Hall-sensor circuits are their low price, their suitability for harsh environments, and their outstanding accuracy due to various error compensating strategies (with zero point errors lower than $0.1$\% of full scale range, and with gain drifts less than $0.5$\% over lifetime and in a wide range of temperatures, $-40\ldots +150$°C). 

This great economical impact has also led to several methods to compute, optimize, and understand the electrical parameters of Hall-plates. Apart from numerical methods like finite element simulation, the most prominent ones are based on conformal transformation -- they were applied to Hall-plates with three or four contacts, symmetric or not, at arbitrary magnetic field  \cite{Wick1954,Lippmann1958,Haeusler1966,Haeusler1967,Versnel1981}. For weak magnetic field and symmetrical Hall-plates conformal transformation was used to express the Hall-output voltage as a function of its resistances between various contacts, without any geometrical parameters  \cite{UPBsimple,ELEN,AusserlechnerSNR2017,Ausserlechner3CHall2018,Ausserlechner2022rspa}. Recently, an analytical method without conformal transformation was proposed \cite{HomentcovschiBercia,Homentcovschi2019}. It works for non-symmetric Hall-plates and arbitrary magnetic field.  Unfortunately, it does not lead to a closed analytical formula for the Hall-output voltage -- instead, it gives two matrices, whose entries are integrals that can be computed only numerically. We will build on these findings in a major part of this work. 

Apart from the direct computation of Hall-plate parameters, one may also ask for relations between a Hall-plate and its complementary Hall-plate, which we get after swapping all contacts and insulating segments of the boundary \cite{Homentcovschi1978,Ausserlechner2019,rspa2,Ausserlechner2022rspa}. These relations are particularly useful, if a Hall-plate is identical to its complementary counterpart, which is the case for strictly regular Hall-plates, the topic of this work. 

The basic task of an engineer of magnetic field sensors is how to increase the magnetic sensitivity, how to reduce the noise, and how to achieve all this with minimum electric power. For a long time it was generally accepted that contacts deteriorate the Hall-output voltage. Therefore, Hall-plates with tiny contacts seemed to be favorable. However, the small contacts increase the noise, and it turned out that traditional Hall-plates with four contacts are noise-optimal if they have medium-sized contacts \cite{AusserlechnerSNR2017}. Besides, Hall-plates with three contacts have less signal-to-noise ratio (SNR) per Watt than Hall-plates with four medium-sized contacts \cite{Ausserlechner3CHall2018,Ausserlechner2016}. This naturally leads to the idea that Hall-plates with more than four contacts might have even better noise performance, and indeed, numerical investigations showed a $90$\% improvement in SNR/sqrt(power) for Hall-plates with 40 contacts \cite{Ausserlechner2020a,Ausserlechner2020hybrid}. However, there are no closed analytical formulae available to compute the parameters of Hall-plates with more than four contacts. This is the basic motivation for the present work.

In Section \ref{sec:fundamentals} we compile some basic definitions, which we will use in this work. Section \ref{sec:RMFoCD-principle} repeats the relations between Hall-plates and complementary Hall-plates. Section \ref{sec:regular-sym} derives properties of Hall-plates with regular symmetry, and Section \ref{sec:strictly-regular-sym} applies all these findings to Hall-plates with strictly regular symmetry. There we compute the eigenvalues of such Hall-plates. In Section \ref{sec:response} we use the eigenvalues to compute the Hall-output voltages for plates with three, four, or infinitely many contacts. Finally, in Section \ref{sec:noise-efficiency} we use the eigenvalues of the Hall-plates to derive the theoretical limits of the thermal noise performance of Hall-plates.

\section{Some basic facts}
\label{sec:fundamentals}

Here we recapitulate some basic properties of Hall-plates, which we will heavily use in the following. In linear electric  networks and in electrically linear Hall-plates the currents into the terminals are linear combinations of the potentials at the terminals. Suppose that the Hall-plate has $N$ terminals. We group all $N$ currents to a vector $\bm{I}$ and all $N$ voltages to a vector $\bm{V}$. Then we express the linear combination as a matrix multiplication $\bm{I} = {^i}\!\bm{G}\bm{V}$, with the \emph{indefinite} conductance matrix ${^i}\!\bm{G}$. Hereby we use the historic nomenclature in circuit theory\cite{Shekel,Haykin,Balabanian,Simonyi}, where 'indefinite' denotes \emph{undefined} reference potential (ground node). In a mathematical sense the indefinite conductance matrix is not indefinite, but positive semi-definite. Indefinite and semi-definite matrices have zero determinants, and therefore ${^i}\!\bm{G}$ cannot be inverted. 
If we ground the $\ell$-th terminal, we write $\bm{I}=\bm{G}\bm{V}$, where we delete the $\ell$-th current and voltage in $\bm{I},\bm{V}$, respectively, and we delete the $\ell$-th row and column in ${^i}\!\bm{G}$ to get the definite conductance matrix $\bm{G}$. 
For any passive (= dissipative) system $\bm{G}$ is positive definite (see Section \ref{sec:strictly-regular-sym}), its determinant is positive, and an inverse exists: $\bm{V}=\bm{R}\bm{I}$ with the resistance matrix $\bm{R}=\bm{G}^{-1}$. 

At zero magnetic field, networks of lumped resistors and transformers are reciprocal, which means that their conductance and resistance matrices are symmetric, $\bm{X}=\bm{X}^T$ for $\bm{X}\in\{\bm{G},\bm{R},{^i}\!\bm{G}\}$ (the superscript ${^T}$ denotes the transpose of a matrix). In the presence of the Hall-effect the entries of the resistance and conductance matrices depend on the magnetic field, $\bm{X}=\bm{X}(\bm{B})$ with the magnetic flux density $\bm{B}$ generated by sources external to the Hall-plate and acting on the plate. Then the matrices $\bm{X}$ are not symmetric, instead Hall-plates are reciprocal only if magnetic field is reversed, 
\begin{equation}\label{eq:fundamentals5}
\left(\bm{X}(\bm{B})\right)^T = \bm{X}(-\bm{B}), \quad \bm{X}\in\{\bm{G},\bm{R},{^i}\!\bm{G}\} .
\end{equation}
This is the principle of \emph{reverse magnetic field reciprocity} (RMFR), which is proven for $\bm{G}$ and $\bm{R}$ in 
Refs.~\cite{Sample1987,Cornils2008}. The validity of (\ref{eq:fundamentals5}) for ${^i}\!\bm{G}$ can be readily proven by its validity for $\bm{G}$. %We note that the RMFR-principle holds also for Hall-plates, if some of their terminals are connected to RMFR-compliant networks (without proof). 
The even parts of the matrices $\bm{X}_\mathrm{ev} = (\bm{X}+\bm{X}^T)/2$ describe again a simple reciprocal network of lumped resistors $r_{k\ell}$ for $k,\ell \in \{0,1,\ldots,N-1\}$; this is the \emph{equivalent resistor circuit} (ERC). Because of the RMFR-principle, $\bm{X}^T=\bm{X}(-\bm{B})$, the matrices $\bm{X}_\mathrm{ev}$ do not depend on the polarity of the applied magnetic field, and consequently the resistors in the ERC depend only on even powers of the magnetic field. Analogously, the odd part $\bm{X}_\mathrm{odd} = (\bm{X}-\bm{X}^T)/2$ depends only on odd powers of the magnetic field. $\bm{X}_\mathrm{odd}$ describes the Hall-effect, whereas $\bm{X}_\mathrm{ev}$ describes the geometrical magneto-resistance effect. (We use the term \emph{geometrical} magneto-resistance to denote the increase in resistances with magnetic field in materials, where the conductance and the Hall-mobility are assumed to stay constant with magnetic field. In real materials, conductance and Hall-mobility also change with the magnetic field, which we call \emph{physical} magneto-resistance effect.)

A singly connected Hall-plate has a single boundary which consists of $N$ contacts $C_\ell$ and $N$ insulating segments $\overline{C}_\ell$. The Hall-effect region is either inside (confined), or outside (extending to infinity). We rule out contacts that are not on the boundary. Then a unique scalar stream function $\psi$ exists \cite{rspa2,{Ausserlechner2019b}}, which describes the current density in the Hall-effect region, 
\begin{equation}\label{eq:RMFoCD1}
\bm{J} = \frac{-1}{\rho} \nabla\times (\psi\bm{n}_z) ,
\end{equation}
with the current density $\bm{J}$, the specific resistivity $\rho$, the nabla operator $\nabla$, the vector product $\times$, and the unit vector $\bm{n}_z$ orthogonal to the plane Hall-plate. In the static case the conservative electric field $\bm{E}$ is given by the negative gradient of the electric potential, $\bm{E}=-\nabla\phi$. In the presence of the Hall-effect Ohm's law is 
\begin{equation}\label{eq:RMFoCD2}
\bm{E}=\rho\bm{J}-\rho\mu_H\bm{J}\times (B_z\bm{n}_z) , 
\end{equation}
with the Hall-mobility $\mu_H$. In (\ref{eq:RMFoCD2}) we use the traditional sign convention for $\mu_H$, which means \emph{negative} Hall-mobility for \emph{electrons} as majority carriers, and we neglect minority carriers. $\bm{E}$ and $\bm{J}$ are not colinear -- there is the Hall-angle $\theta_H$ in-between, with $\tan(\theta_H)=\mu_H B_z$. 
Equation (\ref{eq:RMFoCD2}) links the first partial derivatives of $\phi$ and $\psi$, which are both solutions of the Laplace equation. $\phi$ is homogeneous on the contacts, while $\psi$ is homogeneous on the insulating segments. Conversely, on the insulating segments $\phi$ obeys an oblique derivative boundary condition $ \partial\phi /\partial n=-\mu_H B_z\partial\phi /\partial t$ with $n,t$ denoting the directions normal and tangential to the boundary, while $\psi$ obeys the same boundary condition with opposite sign on the contacts. The opposite sign means a reversal of the magnetic field.

In the sequel we use the following abbreviations, 
\begin{equation}\label{eq:RMFoCD3}
R_\mathrm{sheet} = \frac{\rho}{t_H}, \quad R_\mathrm{sq} = \frac{R_\mathrm{sheet}}{\cos(\theta_H)} ,
\end{equation}
whereby $t_H$ is the thickness of the Hall-plate. If we have a square plate with two contacts fully covering two opposite edges, the resistance between these contacts is the sheet resistance $R_\mathrm{sheet}$ at zero magnetic field, and it is the square resistance $R_\mathrm{sq}$ in the presence of a magnetic field \cite{Lippmann1958}.  

%\section{The $\mathrm{RMFoCD}$-principle}
\section{The RMFoCD-principle}
\label{sec:RMFoCD-principle}

Every Hall-plate has a unique complementary Hall-plate with all contacts and insulating segments swapped. Let us denote all quantities of the complementary Hall-plate with an overbar. We label the contacts of the original Hall-plate $C_0,C_1,\ldots,C_{N-1}$ with increasing index as we walk along the boundaries in positive direction, i.e., with the Hall-effect region being on the left hand side. Thus, for an outer boundary the index increases counter-clockwise, for an inner boundary it increases clockwise. The same applies for the complementary Hall-plate, whereby $\overline{C}_\ell$ is the neighbour of $C_\ell$ in the positive direction. That means, $\overline{C}_\ell$ is in-between $C_\ell$ and $C_{\ell+1}$.

\begin{figure}
  \centering
        \begin{subfigure}[b]{0.47\textwidth}
                \centering
                \includegraphics[width=1.0\textwidth]{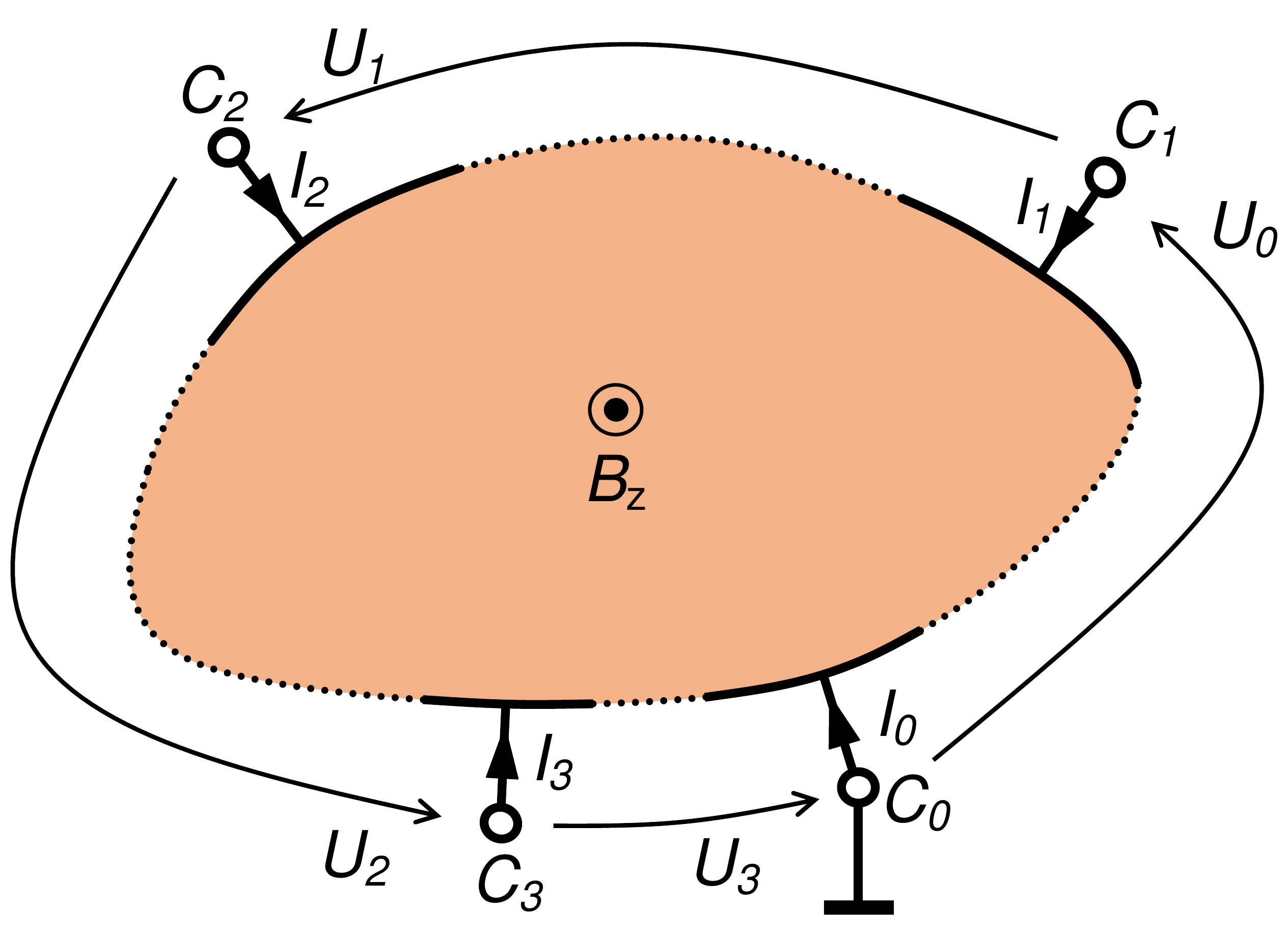}
                \caption{Original Hall-plate at $B_z$.}
                \label{fig:original-case2}
        \end{subfigure}
        \hfill  % An dieser Stelle kann ein zusätzlicher Zwischenraum eingebunden werden: ~, \quad, \qquad, \hfill usw.
          % Eine leere Zeile erzwingt, dass die zweite Grafik darunter erscheint.
        \begin{subfigure}[b]{0.49\textwidth}
                \centering
                \includegraphics[width=1.0\textwidth]{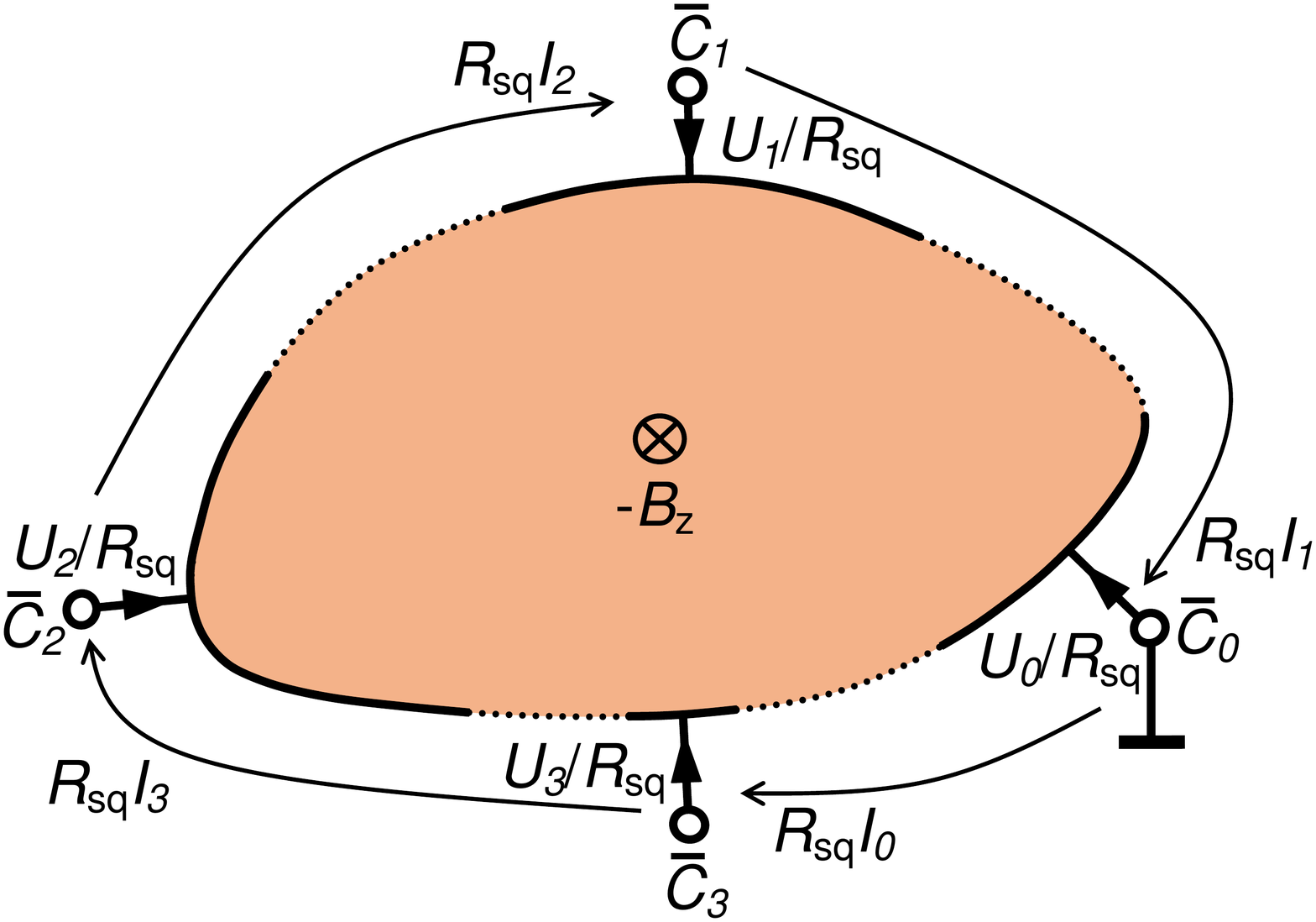}
                \caption{Complementary Hall-plate at $(-B_z)$.}
                \label{fig:complementary-case2}
        \end{subfigure}
    \caption{   RMFoCD principle: An examplary uniform plane Hall-plate has four contacts on the  boundary. It is supplied with currents $I_0,\ldots,I_3$, and it responds with voltages $U_0,\ldots,U_3$ between neighbouring contacts. If the complementary Hall-plate is supplied with currents $\overline{I}_m=U_m/R_\mathrm{sq}$, and if it is exposed to the reverse uniform magnetic field, it responds with voltages $\overline{V}_1-\overline{V}_0=R_\mathrm{sq}I_1$, $\overline{V}_2-\overline{V}_1=R_\mathrm{sq}I_2$, $\overline{V}_3-\overline{V}_2=R_\mathrm{sq}I_3$.} 
   \label{fig:RMFoCD}
\end{figure}

From the complementary boundary conditions for $\phi$ and $\psi$ it follows $\phi = \overline{\psi}$ and $\psi = \overline{\phi}$ at every point inside the Hall-effect region and for any uniform magnetic field, if we inject appropriate supply currents in the original Hall-plate (without overbar) and in the complementary Hall-plate at reverse magnetic field (with overbar). At the contacts this leads to the \emph{Reverse-Magnetic-Field-on-Com-\\plementary-Device} (RMFoCD) principle \cite{Ausserlechner2019,rspa2}.
\begin{equation}\label{eq:RMFoCD5}\begin{split}
\text{For } & \text{arbitrary } I_\ell \text{ it follows } V_\ell \text{ at } \bm{B}. \\
\Rightarrow & \text{ For } \overline{I}_\ell = (V_\ell - V_{\ell+1})/R_\mathrm{sq} \\ 
& \text{ it follows } \overline{V}_\ell - \overline{V}_{\ell-1} = R_\mathrm{sq} I_\ell \text{ at } (-\bm{B}) ,
\end{split}\end{equation}
with $I_\ell, V_\ell$ being the current into $C_\ell$ and the potential of $C_\ell$, respectively. $R_\mathrm{sq}\overline{I}_\ell$ equals the voltage between its neighbouring segments in the original Hall-plate, and $R_\mathrm{sq}I_\ell$ equals the voltage between its neighbouring segments in the complementary Hall-plate (see Fig. \ref{fig:RMFoCD}). 

We can re-write the RMFoCD-principle like this 
\begin{equation}\label{eq:RMFoCD8}\begin{split}
& R_\mathrm{sq} \overline{\bm{I}} = (\bm{1}-\bm{\hat{1}}) \bm{V} \quad\text{and}\quad 
R_\mathrm{sq} \bm{I} = (\bm{1}-\bm{\hat{1}}^T) \overline{\bm{V}} , \text{with} \\
& \bm{1} = 
 \left(\! {\begin{array}{ccccc} 1 & 0 & 0 & \cdots & 0 \\ 0 & 1 & 0 & \cdots & 0 \\ 0 & 0 & 1 & \cdots & 0 \\ \vdots & \vdots & \vdots &\ddots & \vdots \\ 0 & 0 & 0 & \cdots & 1 \\ \end{array} }\!\right) 
\text{ and } \bm{\hat{1}} = 
 \left(\! {\begin{array}{ccccc} 0 & 1 & 0 & \cdots & 0 \\ 0 & 0 & 1 & \cdots & 0 \\ \vdots & \vdots & \vdots &\ddots & \vdots \\ 0 & 0 & 0 & \cdots & 1 \\ 1 & 0 & 0 & \cdots & 0 \\ \end{array} }\!\right) .
\end{split}\end{equation}
$\bm{\hat{1}}$ is a circulant $N\times N$ matrix, which is obtained from the identity matrix $\bm{1}$ by shifting all entries up once, in a rolling way (= by cyclic permutation). It follows $\bm{\hat{1}}^{-1}=\bm{\hat{1}}^T$, $\mathrm{det}(\bm{1}-\bm{\hat{1}})=0$.

If we ground one of the contacts we can express the RMFoCD principle in a third way, 
\begin{equation}\label{eq:RMFoCD9}\begin{split}
& \bm{V} = R_\mathrm{sq} \bm{L_1}^T \overline{\bm{I}} \quad\text{and}\quad \overline{\bm{V}} = R_\mathrm{sq} \bm{L_1} \bm{I} , % \\& 
\quad\text{with} \quad \bm{L_1} = 
 \left(\! {\begin{array}{ccccc} 1 & 0 & 0 & \cdots & 0 \\ 1 & 1 & 0 & \cdots & 0 \\ 1 & 1 & 1 & \cdots & 0 \\ \vdots & \vdots & \vdots &\ddots & \vdots \\ 1 & 1 & 1 & \cdots & 1 \\ \end{array} }\!\right) .
\end{split}\end{equation}
%with $\bm{L_1}$ having entries equal to $0$ above the main diagonal, and all other entries are $1$.
Here, the vectors $\bm{I},\bm{V}$ have only $N-1$ entries and the matrix $\bm{L_1}$ has only $N-1$ rows and columns. The advantage of (\ref{eq:RMFoCD9}) over (\ref{eq:RMFoCD8}) is that $\bm{L_1}$ is invertible, $\mathrm{det}(\bm{L_1})=1$, whereas $\bm{1}-\bm{\hat{1}}$ is singular. Conversely, the advantage of (\ref{eq:RMFoCD8}) over (\ref{eq:RMFoCD9}) is that $\bm{1}-\bm{\hat{1}}$ is a circulant matrix, whereas $\bm{L_1}$ is not.

The resistance matrix of the complementary Hall-plate, $\overline{\bm{R}}$, and the definite conductance matrix of the original Hall-plate, $\bm{G}$, are linked by %(NUMERISCH UEBERPRUEFT)
\begin{equation}\label{eq:RMFoCD30}
\overline{\bm{R}} = R_\mathrm{sq}^2 \bm{L_1} \bm{G}^T \bm{L_1}^T .
\end{equation}
% Formel numerisch geprueft in HomentcovschiMurray_symmetric_irregular_HHalls.nb fuer N=4, unregelmaessige Geometrie bei thetaH=0, 9, 45 degree
This follows from the RMFoCD principle in (\ref{eq:RMFoCD5}); the proof is given in Refs.~\cite{Ausserlechner2019,rspa2}.
In (\ref{eq:RMFoCD30}) all matrices have $(N-1)$ rows and columns, % if the Hall-plates have $N$ contacts. 
and contacts $C_0$ and $\overline{C}_0$ are grounded. 
An alternative form of (\ref{eq:RMFoCD30}) is 
\begin{equation}\label{eq:RMFoCD30b}
\overline{R}_{mn} = R_\mathrm{sq}^2 \sum_{k=1}^{m} \sum_{\ell =1}^{n} G_{\ell k} .
\end{equation}
Thus the entry in the $m$-th row and $n$-th column of the resistance matrix of the complementary Hall-plate is the sum of all entries of the conductance matrix of the original Hall-plate, which are at the intersections of the top $n$ rows with the left $m$ columns, multiplied by $R_\mathrm{sq}^2$.

\section{Regular Hall-plates}
\label{sec:regular-sym}

\begin{figure}
  \centering
                \includegraphics[width=0.53\textwidth]{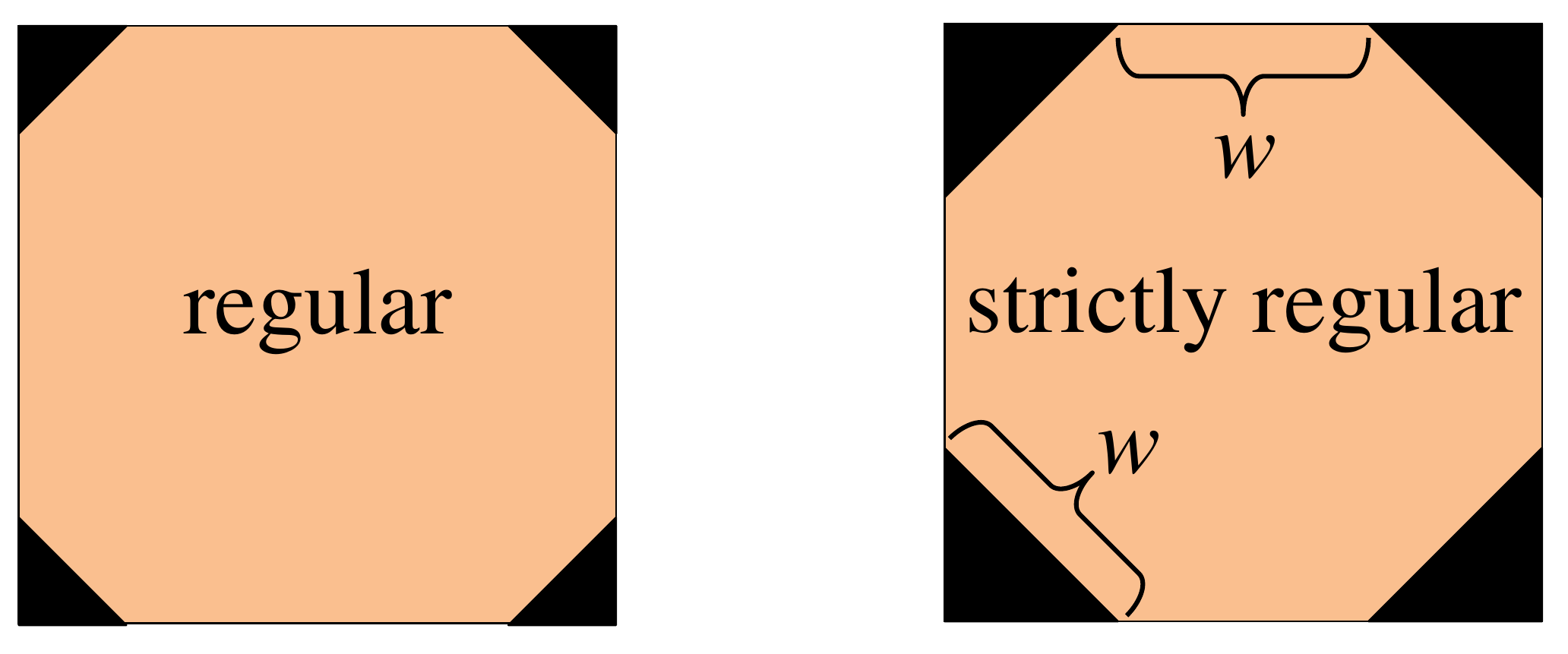}
    \caption{  Examples of regular and strictly regular Hall-plates with four contacts. }
   \label{fig:regular-vs-strictly-regular-Hall-plates}
\end{figure}

Suppose we have a regular boundary with $N$ contacts of equal size, the spacings between the contacts are all equal, yet they are not necessarily identical to the size of the contacts (see Fig. \ref{fig:regular-vs-strictly-regular-Hall-plates}). Let us connect an external circuit with voltage/current sources and volt-/amperemeters to this Hall-plate. If we keep the Hall-plate fixed and move the external circuit by one terminal, say clock-wise, then the resulting potentials and currents into the external circuit must remain the same, because the Hall-plate is regular symmetric in shape and the applied magnetic field is parallel to the rotation axis.This means that the entries in the current and voltage vectors move one instance downwards, and the entries in the indefinite conductance matrix move one row down and one column to the right. Both the original and the new ${^i}\!\bm{G}$ matrix must be identical. We can repeat this procedure $N$ times. The conclusion is that the elements along diagonals parallel to the main diagonal are identical, which gives the following patterns for ${^i}\!\bm{G}$ and $\bm{G}$, 
\begin{equation}\label{G-symmetry}
{^i}\!\bm{G} = \left(\!\!\begin{array}{ccccc} A&B&C&D&E\\ E&A&B&C&D\\ D&E&A&B&C\\ C&D&E&A&B\\ B&C&D&E&A \end{array} \!\!\right) \,\Rightarrow\,
\bm{G} = \left(\!\!\begin{array}{cccc} A&B&C&D\\ E&A&B&C\\ D&E&A&B\\ C&D&E&A \end{array} \!\!\right) .
\end{equation}
${^i}\!\bm{G}$ is an $N\times N$ matrix, $\bm{G}$ is an $(N-1)\times (N-1)$ matrix. We get $\bm{G}$ from ${^i}\!\bm{G}$ by deleting the row and column which corresponds to the grounded contact. $\bm{G}$ and ${^i}\!\bm{G}$ are Toeplitz matrices, which have the property that $G_{\ell ,m}=g_{\ell-m}$ -- it depends only on $\ell-m$. Moreover, ${^i}\!\bm{G}$ is a circulant matrix with $g_{\ell-m}=g_{\ell-m+N}$. Third, the sum of entries per row in any indefinite conductance matrix has to vanish, because no currents flow if all contacts are at identical non-zero potential.

For regular Hall-plates we can get a relation between the \emph{indefinite} conductance matrices of the original and the complementary Hall-plates. With $\bm{I}=\bm{{^i}\!G}\bm{V}$ and $\overline{\bm{I}}=\bm{{^i}}\!\overline{\bm{G}}^T\,\overline{\bm{V}}$ (the transpose accounts for the reverse magnetic field according to the RMFR-principle), and with (\ref{eq:RMFoCD8}) it follows  
\begin{equation}\label{eq:RMFoCD40aa} \begin{split}
(\bm{1}-\bm{\hat{1}}) \bm{V} & = R_\mathrm{sq} \overline{\bm{I}} = R_\mathrm{sq} \,\bm{{^i}}\!\overline{\bm{G}}^T \overline{\bm{V}} \\
\Rightarrow (\bm{1}-\bm{\hat{1}}^T) (\bm{1}-\bm{\hat{1}}) \bm{V} & = R_\mathrm{sq} (\bm{1}-\bm{\hat{1}}^T) \,\bm{{^i}}\!\overline{\bm{G}}^T \overline{\bm{V}} \\
& = R_\mathrm{sq} \,\bm{{^i}}\!\overline{\bm{G}}^T (\bm{1}-\bm{\hat{1}}^T) \overline{\bm{V}} \\
& = R_\mathrm{sq}^2 \,\bm{{^i}}\!\overline{\bm{G}}^T \bm{I} = R_\mathrm{sq}^2 \,\bm{{^i}}\!\overline{\bm{G}}^T \,\bm{{^i}\!G} \bm{V} .
\end{split}\end{equation}
Here we made use of the fact that $\bm{{^i}}\!\overline{\bm{G}}$ and $\bm{1}-\bm{\hat{1}}^T$ are circulant matrices, and therefore they commute. (\ref{eq:RMFoCD40aa}) holds for every $\bm{V}$, therefore
\begin{equation}\label{eq:RMFoCD40a} \begin{split}
2*\bm{1}-\bm{\hat{1}}-\bm{\hat{1}}^T & = R_\mathrm{sq}^2 \,\bm{{^i}}\!\overline{\bm{G}}^T \,\bm{{^i}\!G}, \\ 
2*\bm{1}-\bm{\hat{1}}-\bm{\hat{1}}^T & = R_\mathrm{sq}^2 \,\bm{{^i}}\!\overline{\bm{G}} \,\bm{{^i}\!G}^T .
\end{split}\end{equation}
$2*\bm{1}$ means the multiplication of the scalar number $2$ with the identity matrix $\bm{1}$. The second equation in (\ref{eq:RMFoCD40a}) follows from the first one, if we reverse the magnetic field and apply the RMFR-principle. Alternatively, we can also transpose the first equation and reverse the order of multiplications, because the conductance matrices are circulant and therefore they commute. 

\section{Strictly regular Hall-plates}
\label{sec:strictly-regular-sym}

A Hall-plate is \emph{complementary symmetric} if the complementary plate looks like the original plate (without flipping the plate). Then 50\% of the boundary is covered by contacts. 
\emph{Strictly regular} Hall-plates are both regular and complementary symmetric (see Fig. \ref{fig:regular-vs-strictly-regular-Hall-plates}). When the Hall-plate is mapped to the unit disk, all $N$ contacts and all $N$ insulating segments become equally large -- they subtend the azimuthal angle $\pi/N$. 
This means  
\begin{equation}\label{eq:RMFoCD31-1}
\overline{\bm{R}}=\bm{R} . 
\end{equation}
Inserting this into (\ref{eq:RMFoCD30}) and using $\bm{R}=\bm{G}^{-1}$ gives 
\begin{equation}\label{eq:RMFoCD31}
\bm{1} = R_\mathrm{sq}^2  \bm{G} \bm{L_1} \bm{G}^T \bm{L_1}^T .
\end{equation}
The determinant of both sides of (\ref{eq:RMFoCD31}) is
\begin{equation}\label{eq:RMFoCD32}
\left(\mathrm{det}(R_\mathrm{sq} \bm{G})\right)^2 = 1 \quad\Rightarrow\quad \mathrm{det}(R_\mathrm{sq} \bm{G}) = \pm 1 ,
\end{equation}
because $\mathrm{det}(\bm{L_1})=1$. Next we determine the sign on the right hand side of (\ref{eq:RMFoCD32}).
The power density in the Hall-plate is the dot product of the electric field and the current density. In the presence of the Hall-effect the Hall-angle $\theta_H$ lies between these vectors. This angle varies between $-\pi/2$ and $\pi/2$, i.e., it is acute. Therefore the power density is positive, and also its integral over the Hall-plate volume is positive --- this is the total power $P_d$, . In other words, the Hall-plate is a passive device. On the other hand, the dissipated power in the Hall-plate is  
\begin{equation}\label{eq:RMFoCD33}
P_d = \bm{V}^T \bm{I} = \bm{V}^T \bm{G} \bm{V} > 0 
\end{equation}
for nonvanishing voltage vectors, $\bm{V}\ne\bm{0}$. Therefore, $\bm{G}$ is positive definite, and consequently (i) it is invertible, (ii) all its eigenvalues are positive, and therefore (iii) its determinant is also positive. Using this in (\ref{eq:RMFoCD32}) gives 
\begin{equation}\label{eq:RMFoCD34}
\mathrm{det}(\bm{G}) = R_\mathrm{sq}^{-(N-1)} \quad\Leftrightarrow\quad \mathrm{det}(\bm{R}) = R_\mathrm{sq}^{N-1} ,
\end{equation}
for strictly regular Hall-plates. Inserting (\ref{eq:RMFoCD31-1}) into (\ref{eq:RMFoCD30b}) gives 
\begin{equation}\label{eq:RMFoCD45}
R_{mn} = R_\mathrm{sq}^2 \sum_{k=1}^{m} \sum_{\ell =1}^{n} G_{\ell k} .
\end{equation}
Thus, the entries of the resistance matrix are simple linear combinations of the entries of the conductance matrix (If one applies Cramer's rule for inversion of the $\bm{G}$-matrix, the denominator is simply a power of $R_\mathrm{sq}$). This is a huge simplification over the inversion of general matrices.

For strictly regular Hall-plates it obviously holds $\bm{{^i}}\!\overline{\bm{G}} = \bm{{^i}\!G}$. We use this in (\ref{eq:RMFoCD40a}), whereby we split up the conductance matrix into even and odd portions, $ \bm{{^i}\!G} = \bm{{^i}\!G}_\mathrm{ev}+\bm{{^i}\!G}_\mathrm{odd}$, 
\begin{equation}\label{eq:RMFoCD51a}\begin{split} 
& 2*\bm{1}-\bm{\hat{1}}-\bm{\hat{1}}^T = R_\mathrm{sq}^2 \bm{{^i}\!G}^T \bm{{^i}\!G} \\ 
& \qquad = R_\mathrm{sq}^2 \left(\bm{{^i}\!G}_\mathrm{ev}-\bm{{^i}\!G}_\mathrm{odd} \right) \left(\bm{{^i}\!G}_\mathrm{ev}+\bm{{^i}\!G}_\mathrm{odd} \right) \\ 
& \qquad = R_\mathrm{sq}^2 \left(\bm{{^i}\!G}_\mathrm{ev}^2\underbrace{+\bm{{^i}\!G}_\mathrm{ev}\bm{{^i}\!G}_\mathrm{odd}-\bm{{^i}\!G}_\mathrm{odd}\bm{{^i}\!G}_\mathrm{ev}}_{=\bm{0}} -\bm{{^i}\!G}_\mathrm{odd}^2 \right) ,
\end{split}\end{equation}
whereby we used again the fact that $\bm{{^i}\!G}_\mathrm{ev},\bm{{^i}\!G}_\mathrm{odd}$ are circulant and therefore they commute. Evaluating (\ref{eq:RMFoCD51a}) at zero magnetic field gives 
\begin{equation}\label{eq:RMFoCD51b}
\bm{{^i}}\!G_{\mathrm{ev}0} = \frac{1}{ R_\mathrm{sheet}} \left(2*\bm{1}-\bm{\hat{1}}-\bm{\hat{1}}^T \right)^{1/2} .
\end{equation}
Re-inserting (\ref{eq:RMFoCD51b}) into (\ref{eq:RMFoCD51a}) gives 
\begin{equation}\label{eq:RMFoCD51c}
\bm{{^i}\!G}_\mathrm{odd} = \left( \bm{{^i}\!G}_{\mathrm{ev}}^2 - \left(\cos (\theta_H)\right)^{\!2}\; \bm{{^i}\!G}_{\mathrm{ev}0}^2 \right)^{1/2} .
\end{equation}
For $\theta_H\to\pm\pi/2$ this means $\bm{{^i}\!G}_\mathrm{odd}\to\bm{{^i}\!G}_{\mathrm{ev}}$, which means $\bm{{^i}\!G}_{\mathrm{odd}}\to\bm{0}$ and $\bm{{^i}\!G}_{\mathrm{ev}}\to\bm{0}$, i.e., all conductances vanish.
Equation (\ref{eq:RMFoCD51c}) says that the Hall-response, $\bm{{^i}\!G}_\mathrm{odd}$, is a function of the geometrical magneto-resistance with and without magnetic field, $\bm{{^i}\!G}_\mathrm{ev},\bm{{^i}\!G}_{\mathrm{ev}0}$. Note that the noise of the Hall-plate depends only on the symmetric matrices $\bm{{^i}\!G}_\mathrm{ev},\bm{{^i}\!G}_{\mathrm{ev}0}$, but \emph{not} on the skew-symmetric matrix $\bm{{^i}\!G}_\mathrm{odd}$, because its contribution to the power vanishes due to its skew-symmetry, $P_{d,\mathrm{odd}}=\bm{V}^T\bm{I}_\mathrm{odd}=\bm{V}^T \bm{{^i}\!G}_\mathrm{odd} \bm{V} = 0$ (the quadratic form of any skew-symmetric matrix vanishes). This proves that the Hall-effect portion of the conductance matrix in strictly regular Hall-plates can be measured indirectly via the resistances of the ERC with and without magnetic field \cite{PaulExplicit2009}.

%\subsection{The eigenvalues of $\bm{{^i}\!G}_{\mathrm{ev}0}$ and the zero-field ERC} 
\subsection{The eigenvalues and the ERC at zero-field} 
\label{sec:iGev}

A closed form solution for the diagonalisation of circulant matrices is known \cite{GrayCirculant}.  
\begin{equation}\label{eq:RMFoCD52}\begin{split} 
& 2*\bm{1}-\bm{\hat{1}}-\bm{\hat{1}}^T = \bm{Q} \bm{\Delta} \bm{Q}^C \quad\text{with} \\ 
& (\bm{Q})_{k,\ell} = \frac{1}{\sqrt{N}} \exp\left(\frac{2\pi \mathbbm{i} k \ell }{N}\right)  \;\forall k,\ell\in\{1,2,\ldots,N\} , \\
& \bm{\Delta} = \mathrm{diag}(\Delta_1,\Delta_2,\ldots,\Delta_{N}) , \\ 
& \Delta_\ell = \left(2\sin\left(\frac{\pi\ell}{N}\right)\right)^{\!\!2}, \forall \ell\in\{1,2,\ldots,N\} .
\end{split}\end{equation} 
$\bm{Q}^C$ is the complex conjugate of $\bm{Q}$, and $\mathbbm{i}=\sqrt{-1}$. $\bm{\Delta}$ is a diagonal matrix, which comprises all eigenvalues $\Delta_\ell$ of $2*\bm{1}-\bm{\hat{1}}-\bm{\hat{1}}^T$. It follows $\bm{Q}=\bm{Q}^T$ and $\bm{Q}\bm{Q}^C=\bm{Q}^C\bm{Q}=\bm{1}$. From (\ref{eq:RMFoCD51b}) and (\ref{eq:RMFoCD52}) it follows 
\begin{equation}\label{eq:RMFoCD53}\begin{split}
& \bm{{^i}\!G}_{\mathrm{ev}0} = \bm{Q} \;\bm{{^i}\!\Gamma}_{\mathrm{ev}0} \bm{Q}^C \quad\text{with } \bm{{^i}\!\Gamma}_{\mathrm{ev}0}=\frac{1}{R_\mathrm{sheet}} \sqrt{\bm{\Delta}} \\ 
& \bm{{^i}\!\Gamma}_{\mathrm{ev}0} = \mathrm{diag}({^i}\!\gamma_{\mathrm{ev}0,1},{^i}\!\gamma_{\mathrm{ev}0,2},\ldots,{^i}\!\gamma_{\mathrm{ev}0,N}) , \\
& {^i}\!\gamma_{\mathrm{ev}0,\ell} = \frac{2}{R_\mathrm{sheet}}\sin\left(\frac{\pi\ell}{N}\right), \forall \ell\in\{1,2,\ldots,N\} .
\end{split}\end{equation} 
It means, that the eigenvalues of $\bm{{^i}\!G}_{\mathrm{ev}0}$ are ${^i}\!\gamma_{\mathrm{ev}0,\ell}$. The $N$-th one is zero, all others are positive and smaller than $2/R_\mathrm{sheet}$. This is consistent with the fact that any indefinite conductance matrix must be singular and positive semi-definite. The eigenvectors of $\bm{{^i}\!G}_{\mathrm{ev}0}$ are the columns of $\bm{Q}$, which are coefficient vectors of the discrete Fourier transform (DFT). The explicit values for the conductance matrix at zero magnetic field are 
\begin{equation}\label{eq:RMFoCD55}\begin{split} 
(\bm{{^i}\!G}_{\mathrm{ev}0})_{1,m} & = \frac{2}{N R_\mathrm{sheet}} \!\sum_{\ell=1}^N \sin\!\left(\!\frac{\pi\ell}{N}\!\right) \exp\!\left(\!\frac{-2\pi \mathbbm{i} (m\!-\!1)\ell}{N}\right) \\
& = \frac{1}{R_\mathrm{sheet}}\frac{2}{N}\frac{-\sin(\pi /N)}{\cos(\pi /N)-\cos(2\pi (m-1)/N)} 
\end{split}\end{equation} 
$\forall m\in\{1,2,\ldots,N\} $. Moreover, $(\bm{{^i}\!G}_{\mathrm{ev}0})_{1,m} = -1/r_{1,m}$ for $m\ne 1$ with $r_{1m}$ being the lumped resistors in the ERC. Due to the symmetry of the Hall-plate $r_{1,m}=r_{N,m-1}=r_{m-1,N}$ (whereby contact $C_N$ is identical with contact $C_0$). We finally get 
\begin{equation}\label{eq:RMFoCD56}
r_{m,N} = R_\mathrm{sheet} \frac{N}{2}\; \frac{\cos(\pi /N)-\cos(2\pi m/N)}{\sin(\pi /N)} .
\end{equation} 
Equation (\ref{eq:RMFoCD56}) agrees perfectly with the numerical values reported in Table A1 of Ref.~\cite{Ausserlechner2020a} for $N=3,\ldots, 21$. The smallest resistance in the ERC  at zero magnetic field is 
\begin{equation}\label{eq:RMFoCD57}
\frac{r_{1,N}}{R_\mathrm{sheet}} = \frac{N}{2}\; \frac{\sin(3\pi/(2N))}{\cos(\pi /(2N))} \rightarrow \frac{3\pi}{4} \;\text{ for } N\to\infty .
\end{equation} 
For even $N$, the largest resistance in the ERC for $\bm{B}=\bm{0}$ is 
\begin{equation}\label{eq:RMFoCD58}
\frac{r_{N/2,N}}{R_\mathrm{sheet}} = \frac{N}{2}\; \frac{1+\cos(\pi/N)}{\sin(\pi /N)} \rightarrow \frac{N^2}{\pi} \;\text{ for } N\to\infty .
\end{equation} 
Interestingly, we did not even have to solve the electric field problem of Hall-plates to get the conductance matrix at zero magnetic field. Instead, we only used the RMFoCD principle and the symmetry of the strictly regular Hall-plates. This will change drastically in the next section, where we want to study the magnetic field response of Hall-plates.

%\subsection{The eigenvalues of $\bm{{^i}\!G}$ and the Hall response} 
\subsection{The complex eigenvalues at magnetic field} 
\label{sec:iG}

In (\ref{eq:RMFoCD40a}) all matrices are circulant. Thus, they have identical eigenvectors and therefore identical $\bm{Q}$. This means  
\begin{equation}\label{eq:RMFoCD65}
\bm{{^i}\!G} = \bm{Q} \,\bm{{^i}\!\Gamma} \bm{Q}^C \quad\text{with } \bm{{^i}\!\Gamma} = \mathrm{diag}({^i}\!\gamma_1,{^i}\!\gamma_2,\ldots ,{^i}\!\gamma_N) .
\end{equation} 
Inserting $\bm{{^i}}\!\overline{\bm{G}}=\bm{{^i}\!G}$ into (\ref{eq:RMFoCD40a}) and using (\ref{eq:RMFoCD65}) and (\ref{eq:RMFoCD52}) gives 
\begin{equation}\label{eq:RMFoCD66}\begin{split}
& \bm{Q} \bm{\Delta} \bm{Q}^C = R_\mathrm{sq}^2 \bm{Q}^C \,\bm{{^i}\!\Gamma} \bm{Q}\bm{Q} \,\bm{{^i}\!\Gamma} \bm{Q}^C  \quad\Rightarrow \quad
\bm{Q}\bm{Q} \bm{\Delta} = R_\mathrm{sq}^2 \,\bm{{^i}\!\Gamma} \bm{Q}\bm{Q} \,\bm{{^i}\!\Gamma} , 
\end{split}\end{equation} 
where we multiplied from left and from right with $\bm{Q}$. 
\begin{equation}\label{eq:RMFoCD67}\begin{split}
(\bm{Q}\bm{Q})_{k,m} & = \frac{1}{N} \sum_{\ell=1}^N \exp\left(\frac{2\pi \mathbbm{i} k\ell}{N}\right)\exp\left(\frac{2\pi \mathbbm{i} \ell m}{N}\right) = \delta_{(k+m)\,\mathrm{mod}\,N,0} \\ 
\bm{Q}\bm{Q} & = 
 \left(\! {\begin{array}{cccccc} 0 & 0 & 0 & \cdots & 1 & 0 \\ 0 & 0 & 0 & \cdots & 0 & 0 \\ \vdots & \vdots & \vdots & \reflectbox{$\ddots$} &\vdots & \vdots \\ 0 & 1 & 0 & \cdots & 0 & 0 \\ 1 & 0 & 0 & \cdots & 0 & 0 \\ 0 & 0 & 0 & \cdots & 0 & 1 \\ \end{array} }\!\right) . 
\end{split}\end{equation}
In (\ref{eq:RMFoCD67}) $\delta_{k,\ell}$ is Kronecker's delta, which is $1$ for $k=\ell$ and zero for $k\ne\ell$. Writing out the matrices in (\ref{eq:RMFoCD66}) readily gives  
\begin{equation}\label{eq:RMFoCD68}
R_\mathrm{sq}^2{^i}\!\gamma_\ell {^i}\!\gamma_{N-\ell}=\Delta_\ell \forall\ell\in\{1,2,\ldots ,N-1\} ,\text{ and } {^i}\!\gamma_N = 0 .
\end{equation} 
On the other hand, it generally holds 
\begin{equation}\label{eq:RMFoCD69a}
{^i}\!\gamma_{N-\ell}={^i}\!\gamma_\ell^C ,
\end{equation} 
for the eigenvalues of a real circulant matrix, where $z^C$ is the complex conjugate of $z$. This means 
\begin{equation}\label{eq:RMFoCD69}
|{^i}\!\gamma_\ell |  = \frac{2}{R_\mathrm{sq}}\sin\left(\frac{\pi\ell}{N}\right), \forall \ell\in\{1,2,\ldots,N\} .
\end{equation} 
We use $\bm{{^i}}\!\overline{\bm{G}} = \bm{{^i}\!G} = \bm{{^i}\!G}_\mathrm{ev}+\bm{{^i}\!G}_\mathrm{odd} \Rightarrow \bm{{^i}\!G}^T\bm{{^i}\!G} = \bm{{^i}\!G}_{\bm{\mathrm{ev}}}^2-\bm{{^i}\!G}_{\bm{\mathrm{odd}}}^2$, and insert this into (\ref{eq:RMFoCD40a}). Again, all matrices are circulant with the same $\bm{Q}$ from (\ref{eq:RMFoCD52}). Therefore,  
\begin{equation}\label{eq:RMFoCD70}\begin{split}
& \bm{{^i}\!G}_\mathrm{ev} = \bm{Q} \,\bm{{^i}\!\Gamma_\mathrm{ev}} \bm{Q}^C  \quad\text{and } \bm{{^i}\!G}_\mathrm{odd} = \bm{Q} \,\bm{{^i}\!\Gamma_\mathrm{odd}} \bm{Q}^C , \\
& \bm{{^i}\!\Gamma_\mathrm{ev}} = \mathrm{diag}({^i}\!\gamma_{\mathrm{ev}1},{^i}\!\gamma_{\mathrm{ev}2},\ldots ,{^i}\!\gamma_{\mathrm{ev}N}) , \\
& \bm{{^i}\!\Gamma_\mathrm{odd}} = \mathrm{diag}({^i}\!\gamma_{\mathrm{odd}1},{^i}\!\gamma_{\mathrm{odd}2},\ldots ,{^i}\!\gamma_{\mathrm{odd}N}) , \\
& \Rightarrow \bm{Q} \bm{\Delta} \bm{Q}^C = R_\mathrm{sq}^2 \bm{Q} (\,\bm{{^i}\!\Gamma_\mathrm{ev}}^2 - \,\bm{{^i}\!\Gamma_\mathrm{odd}}^2) \bm{Q}^C .
\end{split}\end{equation}
The eigenvalues of the symmetric matrix $\bm{{^i}\!G}_\mathrm{ev}$ are real, the eigenvalues of the anti-symmetric matrix $\bm{{^i}\!G}_\mathrm{odd}$ are imaginary \cite{GrayCirculant}. From $\bm{{^i}\!G}=\bm{{^i}\!G}_\mathrm{ev}+\bm{{^i}\!G}_\mathrm{odd}$ it follows $\bm{{^i}\!\Gamma}=\bm{{^i}\!\Gamma_\mathrm{ev}}+\bm{{^i}\!\Gamma_\mathrm{odd}}$, which means  
\begin{equation}\label{eq:RMFoCD71}\begin{split}
& {^i}\!\gamma_\ell = {^i}\!\gamma_{\mathrm{ev}\ell} + {^i}\!\gamma_{\mathrm{odd}\ell} \quad \Rightarrow \quad |\,{^i}\!\gamma_\ell |^2  = {^i}\!\gamma_{\mathrm{ev}\ell}^2 + \Im\{\,{^i}\!\gamma_{\mathrm{odd}\ell}\}^2 = {^i}\!\gamma_{\mathrm{ev}\ell}^2 - {^i}\!\gamma_{\mathrm{odd}\ell}^2  ,
\end{split}\end{equation} 
where $\Im\{z\}$ is the imaginary part of the complex number $z$. We apply (\ref{eq:RMFoCD71}) to (\ref{eq:RMFoCD68}) and (\ref{eq:RMFoCD69a}). It gives
\begin{equation}\label{eq:RMFoCD72}\begin{split}
& {^i}\!\gamma_{\mathrm{ev}N-\ell} = {^i}\!\gamma_{\mathrm{ev}\ell}  , \\
& {^i}\!\gamma_{\mathrm{odd}N-\ell} = -{^i}\!\gamma_{\mathrm{odd}\ell}  ,\text{ and } {^i}\!\gamma_{\mathrm{odd}N} = 0 , \\
& \Rightarrow {^i}\!\gamma_{\mathrm{odd}N/2} = 0 \quad \text{for even } N .
\end{split}\end{equation} 

To sum up, the eigenvalues of the indefinite conductance matrix of a strictly regular Hall-plate are real at zero magnetic field (see (\ref{eq:RMFoCD53})). If magnetic field is applied, these eigenvalues rotate from the real axis into the complex plane, whereby their magnitudes decrease (see (\ref{eq:RMFoCD69})), and their real and imaginary portions are the eigenvalues of the even- and odd-symmetric indefinite conductance matrices. The eigenvectors are identical for all these matrices and they are constant versus magnetic field. 

Unfortunately, the RMFoCD-principle does not return the arguments of the complex eigenvalues, which describe the very Hall-effect in the plates. To this end, we need a more specific relation between the potentials and currents in a Hall-plate exposed to a perpendicular magnetic field. For singly-connected domains we can use Ref.~ \cite{Homentcovschi2019}. Their equation (17) reads in our notation
\begin{equation}\label{eq:RMFoCD100}
\sum_{m=1}^N B_{k,m} \phi_m = \sum_{m=1}^N  \frac{-A_{k,m} \psi_m}{\cos(\theta_H)} \text{ for }k\in\{1,\ldots,N-1\} ,
\end{equation} 
whereby $\phi_m$ is the potential on $C_m$ (= the $m$-th peripheral contact), and $\psi_m$ is the stream function on the insulating peripheral segment between $C_m$ and $C_{m+1}$, with $C_N=C_0$. The Hall-plate is mapped to the unit disk, with $C_m$ ranging from azimuthal angle 
\begin{equation}\label{eq:RMFoCD100b}
\alpha_m=2\pi \frac{m-1}{N} \;\text{ to }\; \beta_m=\alpha_m+\frac{\pi}{N} \text{, for } m=1,2,\ldots,N . 
\end{equation} 
In Ref.~\cite{Homentcovschi2019} the matrices $\bm{A},\bm{B}$ are defined as
\begin{equation}\label{eq:RMFoCD101}\begin{split}
& A_{k,m} = \int_{\beta_m}^{\alpha_{m+1}}\frac{h(\tau)\,\mathrm{d}\tau}{\sin((\tau-\beta_k)/2)\sin((\tau-\beta_N)/2)} , \\
& B_{k,m} = \int_{\alpha_m}^{\beta_m}\frac{h(\tau)\,\mathrm{d}\tau}{\sin((\tau-\beta_k)/2)\sin((\tau-\beta_N)/2)}, \\
&\text{with } h(\tau) = \prod_{\ell=1}^N \left| \frac{\sin((\tau-\beta_\ell)/2)}{\sin((\tau-\alpha_\ell)/2)} \right|^{(1/2-\theta_H/\pi)} .
\end{split}\end{equation} 
Note that in (\ref{eq:RMFoCD100}) the sums extend over $N$ terms, whereas in Ref.~\cite{Homentcovschi2019} they had only $N-1$ terms. Hence, in contrast to Ref.~\cite{Homentcovschi2019}, we do not require $\phi_N=0$ and $\psi_N=0$. This enables us to work with the indefinite conductance matrix, which is circulant, and therefore it leads to a simpler mathematical treatment than the definite conductance matrix. Therefore, our matrices $\bm{A},\bm{B}$ have $N$ columns and $N-1$ rows. Equation (\ref{eq:RMFoCD100}) must hold, if we add an arbitrary constant to all $\phi_k$ or to all $\psi_k$. Consequently it must hold 
\begin{equation}\label{eq:RMFoCD102}
\sum_{m=1}^N A_{k,m} = 0 \text{ and } \sum_{m=1}^N B_{k,m} = 0 \text{ for }k\in\{1,\ldots,N-1\} .
\end{equation}
Multiplying the left equation of (\ref{eq:RMFoCD102}) with $\psi_N/\cos(\theta_H)$ and adding it to (\ref{eq:RMFoCD100}) gives 
\begin{equation}\label{eq:RMFoCD103}
\sum_{m=1}^N B_{k,m} \phi_m = \sum_{m=1}^N -A_{k,m} \frac{\psi_m-\psi_N}{\cos(\theta_H)} \text{ for }k\in\{1,\ldots,N-1\} .
\end{equation} 
The currents and the stream function are linked via (8) in Ref.~\cite{Homentcovschi2019} (see also (18) in  Ref.~\cite{Ausserlechner2019b}), %which reads 
\begin{equation}\label{eq:RMFoCD104}
I_k = \frac{\psi_k-\psi_{k-1}}{R_\mathrm{sheet}} .
\end{equation} 
This means 
\begin{equation}\label{eq:RMFoCD105}\begin{split}
& \left.\begin{array}{rcl} I_1&=&(\psi_1-\psi_N)/R_\mathrm{sheet}\\ I_1+I_2&=&(\psi_2-\psi_N)/R_\mathrm{sheet}\\ & \vdots & \\ I_1+\ldots +I_{N-1}&=&(\psi_{N-1}-\psi_N)/R_\mathrm{sheet}\\ I_1+\ldots +I_N &=&0 \end{array} \right\} \\
&\quad\quad \Rightarrow \bm{L_1}\bm{I} = \frac{1}{R_\mathrm{sheet}}\left(\begin{array}{c}\psi_1-\psi_N\\ \psi_2-\psi_N\\ \vdots\\ \psi_{N-1}-\psi_N\\0 \end{array}\right) .
\end{split}\end{equation} 
Inserting (\ref{eq:RMFoCD105}) into (\ref{eq:RMFoCD103}) gives in matrix form 
\begin{equation}\label{eq:RMFoCD106}
\bm{B} \bm{\phi} = -R_\mathrm{sq}\bm{A}\bm{L_1}\bm{I} = -R_\mathrm{sq}\bm{A}\bm{L_1}\bm{{^i}\!G}\bm{\phi} .
\end{equation} 
This equation is valid for arbitrary voltage vectors $\bm{\phi}$, therefore we can skip $\bm{\phi}$ on both sides of (\ref{eq:RMFoCD106}). Multiplying it from right with $(\bm{1}\!-\!\bm{\hat{1}}^T\!)$ gives 
\begin{equation}\label{eq:RMFoCD107}\begin{split}
\bm{B} (\bm{1}\!-\!\bm{\hat{1}}^T\!) & = -R_\mathrm{sq}\bm{A}\bm{L_1}\bm{{^i}\!G}(\bm{1}\!-\!\bm{\hat{1}}^T\!) \\ 
& = -R_\mathrm{sq}\bm{A}\bm{L_1}(\bm{1}\!-\!\bm{\hat{1}}^T\!)\bm{{^i}\!G} ,
\end{split}\end{equation} 
whereby we used the fact that $\bm{{^i}\!G}$ and $(\bm{1}\!-\!\bm{\hat{1}}^T\!)$ are circulant matrices, and therefore they commute. Writing out the product in detail gives $\bm{A}\bm{L_1}(\bm{1}-\bm{\hat{1}}^T) = $ 
\begin{equation}\label{eq:RMFoCD108}\begin{split}
& \bm{A} \left(\! {\begin{array}{ccccc} 1 & 0 & \cdots & 0 \\ 1 & 1 & \ddots & 0 \\ \vdots & \vdots &\ddots & \ddots \\ 1 & 1 & \cdots & 1 \\ \end{array} }\!\right)
\left(\!\begin{array}{ccccc} 1&0&\cdots &0&-1\\ -1&1&\ddots &0&0\\0&-1&\ddots &0 &0\\ \vdots & \ddots & \ddots & 1 & 0 \\ 0&0&\ddots &-1&1 \end{array}\!\right) \\
& = \bm{A}\left(\!\begin{array}{ccccc} 1&0&\cdots &0&-1\\ 0&1&\ddots &0&-1\\0&\ddots &\ddots &\ddots &\vdots \\ 0 & \ddots & \ddots & 1 & -1 \\ 0&0&\cdots &0&0 \end{array}\!\right) = \bm{A} ,
\end{split}\end{equation} 
whereby we used (\ref{eq:RMFoCD102}) in the last equality of (\ref{eq:RMFoCD108}). Equation (\ref{eq:RMFoCD108}) is remarkable, because $\bm{A}\bm{L_1}(\bm{1}-\bm{\hat{1}}^T) = \bm{A}$ even though $\bm{L_1}(\bm{1}-\bm{\hat{1}}^T)\ne\bm{1}$.  Inserting (\ref{eq:RMFoCD108}) into (\ref{eq:RMFoCD107}) gives 
\begin{equation}\label{eq:RMFoCD109}\begin{split}
\bm{B} (\bm{1}\!-\!\bm{\hat{1}}^T\!) & = -R_\mathrm{sq}\bm{A}\bm{{^i}\!G} \\ 
\text{with (\ref{eq:RMFoCD65}) }\;\Rightarrow \bm{B} (\bm{1}\!-\!\bm{\hat{1}}^T\!)\bm{Q} & = -R_\mathrm{sq}\bm{A}\bm{Q} \,\bm{{^i}\!\Gamma} , 
\end{split}\end{equation} 
with 
\begin{equation}\label{eq:RMFoCD110}\begin{split}
& (\bm{B} \bm{\hat{1}}^T \bm{Q})_{j,m} = \sum_{k=1}^N \sum_{\ell=1}^N B_{j,k} \delta_{k,1+(\ell\,\mathrm{mod}\,N)} Q_{\ell ,m} \\ 
& = \sum_{\ell=1}^N B_{j,1+(\ell\,\mathrm{mod}\,N)} Q_{\ell ,m} =  \sum_{k=0}^{N-1} B_{j,k+1} Q_{k,m} . 
\end{split}\end{equation} 
Inserting this into (\ref{eq:RMFoCD109}) gives 
\begin{equation}\label{eq:RMFoCD111}\begin{split}
{^i}\!\gamma_m R_\mathrm{sq} \sum_{k=1}^N A_{j,k} Q_{k,m} = -\sum_{k=1}^N B_{j,k} ( Q_{k,m}-Q_{k-1,m} ) .
\end{split}\end{equation} 
We are not interested in $m=N$, because ${^i}\!\gamma_N=0$. Hence, (\ref{eq:RMFoCD111}) is a solution for the eigenvalues with indices $m=1,2,\ldots,N-1$, whereby (\ref{eq:RMFoCD111}) is valid for any $j\in\{1,2,\ldots,N-1\}$ for each $m$.

\begin{figure*}
  \centering
        \begin{subfigure}[t]{0.95\textwidth}
                \centering
                \includegraphics[width=1.0\textwidth]{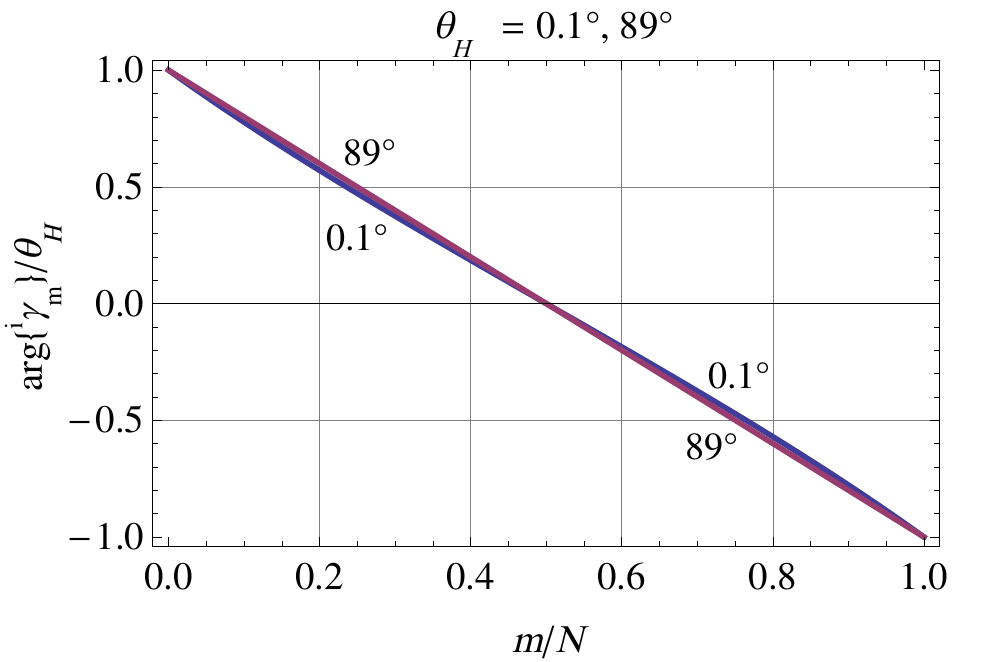}
                \caption{$\mathrm{arg}\{{^i}\!\gamma_m\}/\theta_H$ for $\theta_H=0.1$° and $89$°. }
                \label{fig:argigamma-survey}
        \end{subfigure}
        \\ . \\ . \\ % An dieser Stelle kann ein zusätzlicher Zwischenraum eingebunden werden: ~, \quad, \qquad, \hfill usw.
          % Eine leere Zeile erzwingt, dass die zweite Grafik darunter erscheint.
        \begin{subfigure}[t]{0.95\textwidth}
                \centering
                \includegraphics[width=1.0\textwidth]{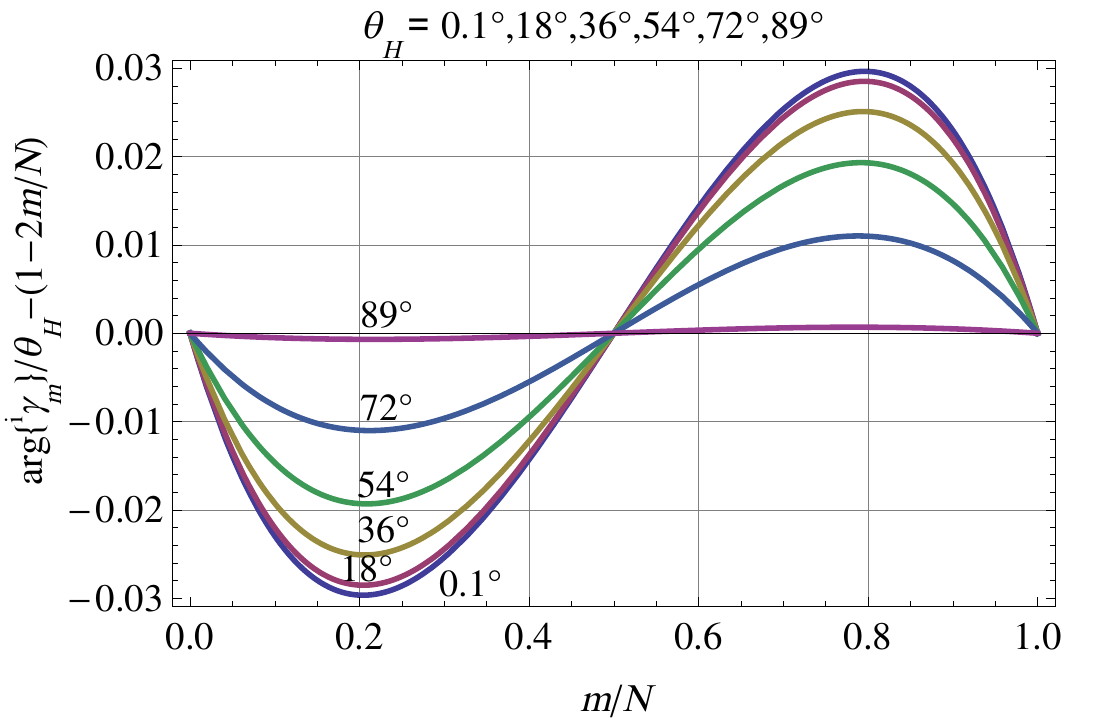}
                \caption{$\mathrm{arg}\{{^i}\!\gamma_m\}/\theta_H\!-\!(1\!-\!2m/N)$ for $\theta_H\!=\!0.1$°, $18$°, $36$°, $54$°, $72$°, $89$°. }
                \label{fig:argigamma-zoom}
        \end{subfigure}
    \caption{Plots of the arguments of the complex eigenvalues ${^i}\!\gamma_m$ of strictly regular Hall-plates versus $m/N$ for various Hall-angles, after (\ref{eq:RMFoCD128}). Fig. \ref{fig:argigamma-survey} gives a general view, Fig. \ref{fig:argigamma-zoom} zooms into the details. }
   \label{fig:argigamma}
\end{figure*}

Inserting (\ref{eq:RMFoCD100b}) into (\ref{eq:RMFoCD101}) and using the product formula \cite{RyshikGradstein} gives 
\begin{equation}\label{eq:RMFoCD112} 
h(\tau) = \left| \cot \left(\frac{N\tau}{2}\right) \right|^{(1/2-\theta_H/\pi)} .
\end{equation} 
Back-inserting this into (\ref{eq:RMFoCD101}) gives 
\begin{equation}\label{eq:RMFoCD113}\begin{split}
& A_{j,k} = \frac{-2}{N} \int_0^{\pi/2}\frac{(\tan(x))^{(1/2-\theta_H/\pi)}\,\mathrm{d}x}{\sin(\frac{x}{N}+\pi \frac{k}{N})\sin(\frac{x}{N} + \pi \frac{k - j}{N})} , \\
& B_{j,k} = \frac{-2}{N} \int_0^{\pi/2}\frac{(\tan(x))^{(1/2-\theta_H/\pi)}\,\mathrm{d}x}{\sin(\frac{x}{N} - \pi \frac{k}{N})\sin(\frac{x}{N} - \pi \frac{k - j}{N})} .
\end{split}\end{equation} 
It follows immediately 
\begin{equation}\label{eq:RMFoCD114}
A_{j,k} = -B_{N-j,N-k} \quad\text{for }\left\{ \begin{array}{l} j\in\{1,2,\ldots N-1\} \\ k\in\{1,2,\ldots N\} \end{array} \right.
\end{equation} 
and 
\begin{equation}\label{eq:RMFoCD115}
\left. \begin{array}{l} A_{j,k+j} = -A_{N-j,k} \\ B_{j,k+j} = -B_{N-j,k} \end{array} \right\} \quad\text{for }\left\{ \begin{array}{l} j\in\{1,2,\ldots N-1\} \\ \text{integer } k \end{array} \right.
\end{equation} 
Inserting $\bm{Q}$ from (\ref{eq:RMFoCD52}) into the right hand side of (\ref{eq:RMFoCD111}) gives 
\begin{equation}\label{eq:RMFoCD116}\begin{split}
& \frac{-1}{\sqrt{N}}\!\left\{\!\left[\sum_{k=1}^{N-1} \!B_{j,k} \exp\!\left(\!\!2\pi \mathbbm{i}\frac{km}{N}\!\right)\!\right] \!\!+\! B_{j,N}\!\right\}\!\!\left(\!1\!-\!\exp\!\left(\!-2\pi \mathbbm{i}\frac{m}{N} \right)\!\right) \\
& =\frac{2\mathbbm{i}}{\sqrt{N}}\exp\left(-\mathbbm{i}\pi\frac{m}{N}\right) \sin\left(\pi\frac{m}{N}\right) \sum_{k=1}^{N-1} A_{N-j,N-k} \left[\exp\left(2\pi \mathbbm{i}\frac{km}{N}\right)\!-\!1 \right] ,
\end{split}\end{equation} 
where we used (\ref{eq:RMFoCD102}) and (\ref{eq:RMFoCD114}). With $\ell=N-k$ this gives 
\begin{equation}\label{eq:RMFoCD117}\begin{split}
& \frac{2\mathbbm{i}}{\sqrt{N}}\exp\!\left(\!-\!\mathbbm{i}\pi\frac{m}{N}\right) \!\sin\!\left(\!\pi\frac{m}{N}\!\right)\! \sum_{\ell=1}^{N-1} \!A_{N-j,\ell} \left[\exp\!\left(\!-\!2\pi \mathbbm{i}\frac{\ell m}{N}\!\right)\!-\!1 \right] \\ 
& = \frac{2\mathbbm{i}}{\sqrt{N}}\exp\!\left(\!-\!\mathbbm{i}\pi\frac{m}{N}\right) \!\sin\!\left(\pi\frac{m}{N}\!\right)\! \sum_{k=0}^{N-1} \!A_{N-j,k} \exp\!\left(\!-\!2\pi \mathbbm{i}\frac{km}{N}\!\right) .
\end{split}\end{equation} 
With $A_{N-j,k} = -A_{j,(k+j)\,\mathrm{mod}\,N}$ and the replacement $\ell=k+j$ the sum in (\ref{eq:RMFoCD117}) becomes 
\begin{equation}\label{eq:RMFoCD118}\begin{split}
&  \sum_{\ell=j}^{N+j-1} -A_{j,\ell\,\mathrm{mod}\,N} \exp\left(-2\pi \mathbbm{i} m\frac{\ell-j}{N}\right) \\
& \quad = \sum_{\ell=j}^{N-1} -A_{j,\ell}\exp\left(-2\pi \mathbbm{i}\frac{m\ell}{N}\right) \exp\left(2\pi \mathbbm{i}\frac{mj}{N}\right) \\ 
& \quad + \sum_{\ell=0}^{j-1} -A_{j,\ell}\exp\left(-2\pi \mathbbm{i}\frac{m\ell}{N}\right) \exp\left(2\pi \mathbbm{i}\frac{mj}{N}\right) .
\end{split}\end{equation} 
Using this in (\ref{eq:RMFoCD111}) gives 
\begin{equation}\label{eq:RMFoCD119}\begin{split}
& {^i}\!\gamma_m R_\mathrm{sq} \sum_{k=1}^N A_{j,k} \exp\left(2\pi \mathbbm{i}\frac{km}{N}\right) \\ 
& = -2 \sin\left(\pi\frac{m}{N}\right) \mathbbm{i}\exp\left(\mathbbm{i}\pi m\frac{2j-1}{N}\right) \sum_{k=1}^N A_{j,k} \exp\left(-2\pi \mathbbm{i}\frac{km}{N}\right) .
\end{split}\end{equation} 
The sums on both sides of this equation are complex conjugate numbers. Hence, it follows immediately (\ref{eq:RMFoCD69}) for the magnitudes of the eigenvalues ${^i}\!\gamma_m$. In the Appendix we prove the identity 
\begin{equation}\label{eq:RMFoCD125}\begin{split}
& \sum_{k=1}^N \frac{\exp\left(2\pi \mathbbm{i} k m / N\right)}{\sin(x/N\!+\!\pi k/N)\sin(x/N\!+\!\pi (k\!-\!j)/N)} \\
& = 2N\frac{\sin\left(\pi j m / N\right)}{\sin\left(\pi j / N\right)} \left(\mathbbm{i}\cot(x)\!-\!1\right)\exp\left(\mathbbm{i}m \frac{j\pi\!-\!2x}{N}\right) .
\end{split}\end{equation} 
This sum vanishes if the product $j m$ is an integer multiple of $N$. We have to avoid such values of $j$, because they make both sides of (\ref{eq:RMFoCD111}) equal to zero, and this annihilates any information on ${^i}\!\gamma_m$.  
Using (\ref{eq:RMFoCD125}) and (\ref{eq:RMFoCD113}) in (\ref{eq:RMFoCD119}) gives the arguments of the complex eigenvalues 
\begin{equation}\label{eq:RMFoCD126}\begin{split}
& \mathrm{arg}\{{^i}\!\gamma_m\} = \frac{-\pi}{2}+2\,\mathrm{arg}\biggl\{ \\
& \left.\quad\int_0^{\pi/2} \left(\tan(x)\right)^{-\theta_H/\pi} \left(\sqrt{\tan(x)}+\mathbbm{i}\sqrt{\cot(x)}\right) \exp\left(\mathbbm{i}\frac{m}{N}\left(2x\!-\!\frac{\pi}{2}\right)\right) \mathrm{d}x \right\} .
\end{split}\end{equation} 
% Diese Formel habe ich numerisch verifiziert gegen Homentcovschi und Murray fuer n=3,4,5,6,7,12,19,20,22 fuer thetaH = 1, 10, 45 degree.
In this equation $j$ has cancelled out. Moreover, the eigenvalues depend only on the ratio $m/N$, not on the absolute values of $m,N$. We use the identity 
\begin{equation}\label{eq:RMFoCD127}
\sqrt{\tan(x)}+\mathbbm{i}\sqrt{\cot(x)} = \sqrt{\frac{2}{\sin(2x)}} \exp\left(\mathbbm{i}\left(\frac{\pi}{2}-x\right)\right) ,
\end{equation} 
and we split up the integration domain into $0\le x\le\pi/4$ and $\pi/4\le x\le\pi/2$. The final result is 
\begin{equation}\label{eq:RMFoCD128}\begin{split}
& \mathrm{arg}\{{^i}\!\gamma_m\} = 2\,\mathrm{arg} \biggl\{ \\
& \qquad\quad\int_0^{\pi/4}\!\!\frac{1}{\sqrt{\sin(2x)}} \left[ \left(\tan(x)\right)^{-\theta_H/\pi} \exp\left(\!-\mathbbm{i}\left(x\!-\!\frac{\pi}{4}\right) \left(1\!-\!\frac{2m}{N}\right)\right) \right. \\
& \qquad\qquad\qquad\qquad\quad\left.\left. +\left(\tan(x)\right)^{\theta_H/\pi} \exp\left(\mathbbm{i}\left(x\!-\!\frac{\pi}{4}\right) \left(1\!-\!\frac{2m}{N}\right)\right) \right]  \mathrm{d}x \right\} .
\end{split}\end{equation}
From this equation we see immediately, that the arguments of all eigenvalues are \emph{odd} functions of the Hall-angle. This implies that the eigenvalues become real at zero Hall-angle. (\ref{eq:RMFoCD69a}) also follows from (\ref{eq:RMFoCD128}). In the limit of very large Hall-angles it holds 
\begin{equation}\label{eq:RMFoCD129}\begin{split}
 \lim_{\theta_H\to\pi/2}\mathrm{arg}\{{^i}\!\gamma_m\} & = 2\,\mathrm{arg}\left\{ \int_0^{\pi/4} \frac{\exp\left(-\mathbbm{i}\left(x-\pi/4\right) \left(1-2m/N\right)\right)}{\sin(x)} \mathrm{d}x \right\} \\ 
& = \frac{\pi}{2}-\pi\frac{m}{N} ,
\end{split}\end{equation}
because the integral over the second summand in the integrand of (\ref{eq:RMFoCD128}) remains finite while the integral over the first summand grows unboundedly. Hence, at infinite magnetic field the arguments of the eigenvalues are uniformly distributed between $-90$° and $90$°. 
The limit of weak magnetic field has much more practical relevance. There we replace $(\tan(x))^{\pm\theta_H/\pi}\to 1\pm(\theta_H/\pi)\ln(\tan(x))$ and we get 
\begin{equation}\label{eq:RMFoCD130}\begin{split}
 \lim_{\theta_H\to 0}\frac{\mathrm{arg}\{{^i}\!\gamma_m\}}{\theta_H} = \frac{\int_0^{\pi/4} \;\ln\left(\tan(x)\right) \frac{\sin\left(\left(x-\pi/4\right) \left(1-2m/N\right)\right)}{\sqrt{\sin(2x)}} \mathrm{d}x}{\frac{\pi}{2} \int_0^{\pi/4} \;\frac{\cos\left(\left(x-\pi/4\right) \left(1-2m/N\right)\right)}{\sqrt{\sin(2x)}} \mathrm{d}x} .
\end{split}\end{equation}
Hence, at weak magnetic field the arguments of the eigenvalues are linearly proportional to the Hall-angle. Although ${^i}\!\gamma_0$ does not exist, we are still allowed to set $m=0$ in (\ref{eq:RMFoCD130}). Then the right hand side becomes $1$ (Both the numerator and the denominator are integrable in closed form, e.g. with the computational software program Wolfram Mathematica. They are equal to $\pi^2/(4\sqrt{2})$).
A numerical inspection of (\ref{eq:RMFoCD130}) (e.g. with the standard procedure {\tt NIntegrate} of Mathematica) suggests the following series expansion 
\begin{equation}\label{eq:RMFoCD130b}\begin{split}
\lim_{\theta_H\to 0}\frac{\mathrm{arg}\{{^i}\!\gamma_m\}}{\theta_H} = 1-2\frac{m}{N}+\sum_{k=0}^\infty a_k \sin\left(2\pi k\frac{m}{N}\right) .
\end{split}\end{equation}
The coefficients are obtained from a numerical fit of (\ref{eq:RMFoCD130}),
\begin{equation}\label{eq:RMFoCD130c}\begin{split}
a_1 = 0.0297118586, & \quad a_2 = 0.004343366943, \\
a_3 = 0.00133654978, & \quad a_4 = 0.000572028149, \\
a_5 = 0.000294907531, & \quad a_6 = 0.0001713175541 .
\end{split}\end{equation}
In (\ref{eq:RMFoCD130b}) the term $1-2m/N$ dominates and the infinite sum is only a small correction, which is hardly visible by the naked eye when its graph is plotted in Fig. \ref{fig:argigamma}.
With a numerical evaluation of  (\ref{eq:RMFoCD130}) and with (\ref{eq:RMFoCD129}) it follows 
\begin{equation}\label{eq:RMFoCD131}\begin{split}
& N=3:\; \mathrm{arg}\{{^i}\!\gamma_1\} = \left(0.311078705\ldots \frac{1}{3}\right)\theta_H , \\
& N=4:\; \mathrm{arg}\{{^i}\!\gamma_1\} = \left(\frac{\sqrt{2}}{3}\ldots \frac{1}{2}\right)\theta_H , \\
& N=\infty:\; \mathrm{arg}\{{^i}\!\gamma_1\} = \left(1\ldots 1\right)\theta_H ,
\end{split}\end{equation}
whereby the left numbers denote the weak magnetic field limit and the right numbers denote the strong magnetic field limit. Hence, for strictly regular Hall-plates with many contacts $\mathrm{arg}\{{^i}\!\gamma_1\}/\theta_H$ does not depend on the magnetic field. This is plausible, because for $N\to\infty$ the contacts tend to be point-sized, and for point-sized contacts the Hall-voltages (and therefore also the arguments of the eigenalues) are linear functions of the Hall-angle, as it was shown in Ref.~\cite{Ausserlechner2019b}. Converseley, the strongest field dependence of $\mathrm{arg}\{{^i}\!\gamma_1\}/\theta_H$ occurs for lowest contact count, $N=3$. There it varies by $7.2$\% for $\theta_H=0\to 90$°. 

% original file: noise-eficiency_of_strictly_regular_Halls.nb

\section{Magnetic field response for strictly regular Hall-plates}
\label{sec:response}
We apply these findings to the cases $N=3,4,\infty$ to get the conductance matrix and the Hall-geometry factor. 

\subsection{Strictly regular Hall-plates with three contacts}
\label{sec:3CHall}
We compute $\bm{Q}$ from (\ref{eq:RMFoCD52}) and insert it into (\ref{eq:RMFoCD65}) to get the conductance matrix. 
\begin{equation}\label{eq:RMFoCD150}\begin{split}
G_{1,1} & = \frac{\cos(\theta_H)}{R_\mathrm{sheet}}\frac{2}{\sqrt{3}}\cos\left(\mathrm{arg}\{{^i}\!\gamma_1\}\right) , \\
G_{1,2} & = \frac{\cos(\theta_H)}{R_\mathrm{sheet}}\frac{2}{\sqrt{3}}\cos\left(\mathrm{arg}\{{^i}\!\gamma_1\}-\frac{2\pi}{3}\right) , \\
G_{2,1} & = \frac{\cos(\theta_H)}{R_\mathrm{sheet}}\frac{2}{\sqrt{3}}\cos\left(\mathrm{arg}\{{^i}\!\gamma_1\}+\frac{2\pi}{3}\right) .
\end{split}\end{equation}
All other entries of the conductance matrix readily follow from the fact that it is a circulant matrix. In Ref.~\cite{Ausserlechner2016} we defined the Hall-geometry factor for a three-contacts Hall-plate, $G_H^{(3C)}$, in the following way: contact $C_0$ is grounded, supply current $I_\mathrm{supply}$ is injected into contact $C_1$, and the potential $V_2$ is measured at contact $C_2$. Then $G_H^{(3C)}$ is given by 
\begin{equation}\label{eq:RMFoCD151}
V_2(\theta_H)-V_2(-\theta_H) = R_\mathrm{sheet}\tan(\theta_H) I_\mathrm{supply} G_H^{(3C)} .
\end{equation}
This definition implies that $G_H^{(3C)}\to 1$ for point-sized contacts, while $G_H^{(3C)}\to 0$ when the contacts spacings vanish. We use (\ref{eq:RMFoCD150}) to compute the left hand side in (\ref{eq:RMFoCD151}). It gives 
\begin{equation}\label{eq:RMFoCD152}
G_H^{(3C)} = 2\frac{\sin\left(\mathrm{arg}\{{^i}\!\gamma_1\}\right)}{\sin(\theta_H)}
\end{equation}
Inserting (\ref{eq:RMFoCD131}) into (\ref{eq:RMFoCD152}) gives the limits for weak and strong magnetic field  
\begin{equation}\label{eq:RMFoCD153}\begin{split}
\lim_{\theta_H\to 0} G_H^{(3C)} & = G_{H0}^{(3C)} = 0.6221574108 , \\
\lim_{\theta_H\to \pi/2} G_H^{(3C)} & = G_{H\infty}^{(3C)} = 1 . 
\end{split}\end{equation}
The strong magnetic field limit is plausible, because at $\theta_H\to 90$° Hall-plates with extended contacts behave like Hall-plates with point-sized contacts (see Section 9 in Ref.~\cite{Ausserlechner2019b}). The weak magnetic field limit in (\ref{eq:RMFoCD153}) is identical to the numerical value reported in Ref.~\cite{Ausserlechner3CHall2018}. A plot of (\ref{eq:RMFoCD152}) with ${^i}\!\gamma_1$ from (\ref{eq:RMFoCD128}) is identical to Figure 17 in Ref.~\cite{Ausserlechner3CHall2018}. A comparison of (\ref{eq:RMFoCD153}) with (\ref{eq:RMFoCD131}) is also interesting: $\mathrm{arg}\{{^i}\!\gamma_1\}/\theta_H$ changes only $7$\% for $\theta_H=0\to 90$°, yet $G_H^{(3C)}$ changes much more, namely $48$\%. 

In (\ref{eq:RMFoCD152}) we use $\mathrm{arg}\{{^i}\!\gamma_1\}$ from (\ref{eq:RMFoCD128}) with $N=3$. Numerical inspection shows that $\mathrm{arg}\{{^i}\!\gamma_1\}/\theta_H$ is close to a parabola versus the Hall-angle. This gives a good approximation of the Hall-geometry factor 
\begin{equation}\label{eq:RMFoCD154}\begin{split}
G_H^{(3C)} \approx & \frac{2}{\sin(\theta_H)} \sin\!\left(\!\!\theta_H\!\left[\!\frac{G_{H0}^{(3C)}}{2}\!+\!\left(\!\frac{1}{3}\!-\!\frac{G_{H0}^{(3C)}}{2}\right)\!\left(\!\frac{\theta_H}{\pi/2}\!\right)^{\!\!2} \right] \right) 
\end{split}\end{equation}
% in noise-efficiency_of_strictly_regular_Halls.nb
The relative error of (\ref{eq:RMFoCD154}) in $-\pi/2\le\theta_H\le\pi/2$ is $+0.153$\%,$-0$\%. The error vanishes at $\theta_H=0,\pm \pi/2$.

\subsection{Strictly regular Hall-plates with four contacts}
\label{sec:4CHall}

These types of Hall-plates are of paramount practical relevance. Exact formulae from Haeusler in Ref.~\cite{Haeusler1966} used the hypergeometric and the beta function. Our approach uses the complex-valued integral (\ref{eq:RMFoCD128}) instead. The conductance matrix follows from (\ref{eq:RMFoCD65}), 
\begin{equation}\label{eq:RMFoCD160}\begin{split}
G_{1,1} & = \frac{\cos(\theta_H)}{R_\mathrm{sheet}}\frac{1}{2}\left(1+\sqrt{2}\cos\left(\mathrm{arg}\{{^i}\!\gamma_1\}\right)\right) , \\
G_{1,2} & =\frac{\cos(\theta_H)}{R_\mathrm{sheet}}\frac{1}{2}\left(-1+\sqrt{2}\sin\left(\mathrm{arg}\{{^i}\!\gamma_1\}\right)\right) , \\
G_{1,3} & = \frac{\cos(\theta_H)}{R_\mathrm{sheet}}\frac{1}{2}\left(1-\sqrt{2}\cos\left(\mathrm{arg}\{{^i}\!\gamma_1\}\right)\right) .
\end{split}\end{equation}
$G_{2,1}$ is obtained from $G_{1,2}$ by multiplying $\mathrm{arg}\{{^i}\!\gamma_1\}$ with $-1$. The second eigenvalue is real, ${^i}\!\gamma_2=2/R_\mathrm{sq}$, its argument vanishes. The Hall-geometry factor is commonly defined by 
\begin{equation}\label{eq:RMFoCD161}\begin{split}
V_3-V_1 & = V_3(\theta_H)-V_3(-\theta_H) = V_1(-\theta_H)-V_1(\theta_H) \\ 
& = R_\mathrm{sheet}\tan(\theta_H) I_\mathrm{supply} G_H^{(4C)} ,
\end{split}\end{equation}
whereby $I_\mathrm{supply}$ flows into contact $C_2$ through the Hall-plate towards the opposite grounded contact $C_0$, and a voltmeter samples the voltage across the output contacts $C_3-C_1$. In (\ref{eq:RMFoCD161}) $V_k$ means $V_k(\theta_H)$ for $k=1,3$. 
$G_H^{(4C)}\to 1$ for point-sized contacts, while $G_H^{(4C)}\to 0$ if the insulating segments between the contacts vanish. Applying (\ref{eq:RMFoCD160}) to compute the left hand side in (\ref{eq:RMFoCD161}) gives 
\begin{equation}\label{eq:RMFoCD162}
G_H^{(4C)} = \sqrt{2}\frac{\sin\left(\mathrm{arg}\{{^i}\!\gamma_1\}\right)}{\sin(\theta_H)}
\end{equation}
Inserting (\ref{eq:RMFoCD131}) into (\ref{eq:RMFoCD162}) gives the limits for weak and strong magnetic field  
\begin{equation}\label{eq:RMFoCD163}\begin{split}
\lim_{\theta_H\to 0} G_H^{(4C)} & = G_{H0}^{(4C)} = 2/3 , \\
\lim_{\theta_H\to \pi/2} G_H^{(4C)} & = G_{H\infty}^{(4C)} = 1 . 
\end{split}\end{equation}
The strong magnetic field limit is plausible for the same reason as in (\ref{eq:RMFoCD153}). A plot of (\ref{eq:RMFoCD162}) with ${^i}\!\gamma_1$ from (\ref{eq:RMFoCD128}) is identical to Figure 4 in Ref.~\cite{AusserlechnerSNR2017}. 
The weak magnetic field limit in (\ref{eq:RMFoCD163}) is identical to the value reported in Ref.~\cite{Haeusler1967}. 

In (\ref{eq:RMFoCD162}) we use $\mathrm{arg}\{{^i}\!\gamma_1\}$ from (\ref{eq:RMFoCD128}) with $N=4$. Numerical inspection shows that $\mathrm{arg}\{{^i}\!\gamma_1\}/\theta_H$ is close to a parabola versus the Hall-angle. This gives a good approximation of the Hall-geometry factor 
\begin{equation}\label{eq:RMFoCD164}
G_H^{(4C)} \!\approx \!\frac{\sqrt{2}}{\sin(\theta_H)} \!\sin\!\left(\!\!\theta_H\!\left[\!\frac{\sqrt{2}}{3}\!+\!\left(\!\frac{1}{2}\!-\!\frac{\sqrt{2}}{3}\right)\!\left(\!\frac{\theta_H}{\pi/2}\!\right)^{\!\!2} \right] \right) 
\end{equation}
% in noise-efficiency_of_strictly_regular_Halls.nb
The relative error of (\ref{eq:RMFoCD164}) in $-\pi/2\le\theta_H\le\pi/2$ is $+0.09$\%,$-0$\%. The error vanishes at $\theta_H=0,\pm \pi/2$. 

%\subsection{Strictly regular Hall-plates with $N$ contacts}
\subsection{Strictly regular Hall-plates with N contacts}
\label{sec:NCHall}

\begin{figure}
  \centering
                \includegraphics[width=0.90\textwidth]{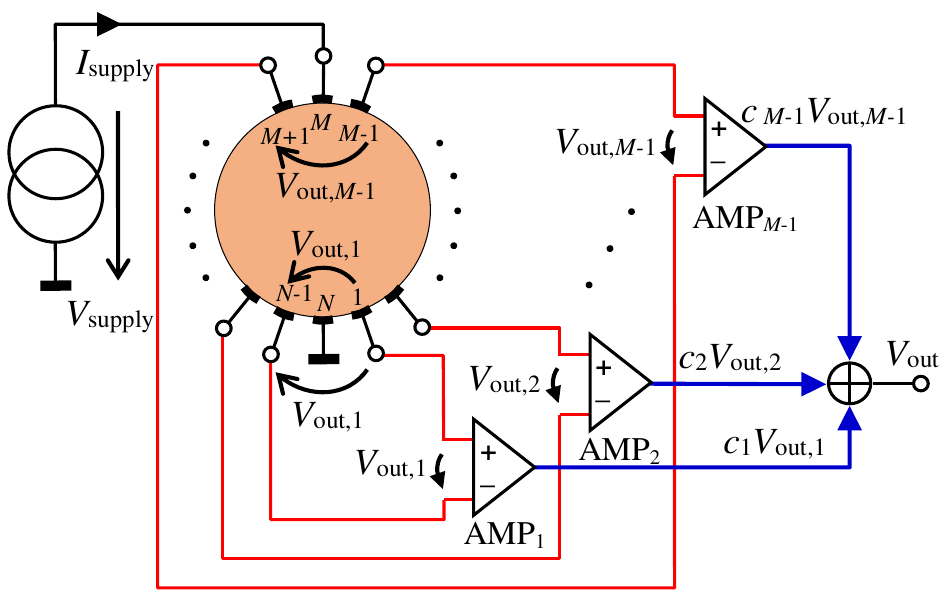}
    \caption{  Hall-plate with $N=2M$ contacts with a single supply current and $M-1$ output voltages, which are summed up with weighting coefficients $c_k$ by $M-1$ amplifiers $\mathrm{AMP}_k$. (Figure adapted from Ref.~\cite{Ausserlechner2020a}.) }
   \label{fig:single-input-current-multiple-output-voltages}
\end{figure}

For an arbitrary number of contacts we use (\ref{eq:RMFoCD65}), (\ref{eq:RMFoCD68}), (\ref{eq:RMFoCD69}) and (\ref{eq:RMFoCD128}). 
\begin{equation}\label{eq:RMFoCD170}\begin{split}
G_{\ell ,k} & = (-1)^{\ell-k}\frac{4}{N R_\mathrm{sq}} \left(\frac{N-1}{2}-\left\lfloor\frac{N}{2}\right\rfloor \right) \\
& \quad + \frac{4}{N R_\mathrm{sq}} \sum_{p=1}^{\left\lfloor N/2\right\rfloor} \!\sin\!\left(\!\pi\frac{p}{N}\right) \cos\!\left( \!\mathrm{arg}\{{^i}\!\gamma_p\} \!+\! 2\pi p\frac{\ell\!-\!k}{N} \right) ,
\end{split}\end{equation} 
whereby $\lfloor M\rfloor$ is the integer part of the positive real number $M$. In (\ref{eq:RMFoCD170}) it is easy to get the even part of the conductance matrix, $(\bm{{^i}G_\mathrm{ev}})_{\ell ,k}$ (=magneto-resistive term), and the odd part, $(\bm{{^i}G_\mathrm{odd}})_{\ell ,k}$ (=Hall term), simply by splitting up the cosine-term with the goniometric formula.
Inserting (\ref{eq:RMFoCD170}) into (\ref{eq:RMFoCD45}) gives the resistance matrix 
\begin{equation}\label{eq:RMFoCD171}\begin{split}
R_{\ell ,k} & = R_\mathrm{sq} \frac{16}{N}\! \left(\!\frac{\ell}{2}\!-\!\left\lfloor\frac{\ell}{2}\right\rfloor\!\right)\! \left(\!\frac{k}{2}\!-\!\left\lfloor\frac{k}{2}\right\rfloor\!\right)\! \left(\frac{N\!-\!1}{2}\!-\!\left\lfloor\frac{N}{2}\right\rfloor\right)  \\
&\quad+ R_\mathrm{sq}\frac{4}{N} \sum_{p=1}^{\left\lfloor N/2\right\rfloor} \frac{\sin\left(\pi p\ell /N\right)\sin\left(\pi p k/N\right)}{\sin\left(\pi p/N\right)} \cos\left( \mathrm{arg}\{{^i}\!\gamma_p\}- \pi p\frac{\ell-k}{N} \right) .
\end{split}\end{equation}
\noindent Now we are able to compute the Hall-output voltages of a strictly regular plate with $N=2M$ contacts, where current $I_\mathrm{supply}$ is injected into contact $C_M$ and contact $C_N$ is grounded, as shown in Fig. \ref{fig:single-input-current-multiple-output-voltages}. $M-1$ output voltages $V_{\mathrm{out},k}$ are tapped between $C_k-C_{N-k}$. Analogous to (\ref{eq:RMFoCD161}) we define the Hall geometry factor $G_{H,k}^{(2M\times C)}$ 
\begin{equation}\label{eq:RMFoCD175}
V_{\mathrm{out},k} = -R_\mathrm{sheet} \tan{\theta_H} I_\mathrm{supply} G_{H,k}^{(2M\times C)} .
\end{equation}
The minus sign means that $V_{\mathrm{out},k}<0$ for $\theta_H>0$. This gives 
\begin{equation}\label{eq:RMFoCD176}\begin{split}
G_{H,k}^{(2M\times C)} & = \frac{R_{N-k,M}-R_{k,M}}{R_\mathrm{sheet} \tan{\theta_H}} = \frac{2}{M}  \sum_{m=1,3,5\ldots}^{M-1} \frac{\sin\left(\pi\frac{km}{M}\right)}{\sin\left(\pi\frac{m}{2M}\right)} \;\frac{\sin\left(\mathrm{arg}\{{^i}\!\gamma_m\}\right)}{\sin\left(\theta_H\right)}  .
\end{split}\end{equation}

\begin{figure}
  \centering
                \includegraphics[width=0.85\textwidth]{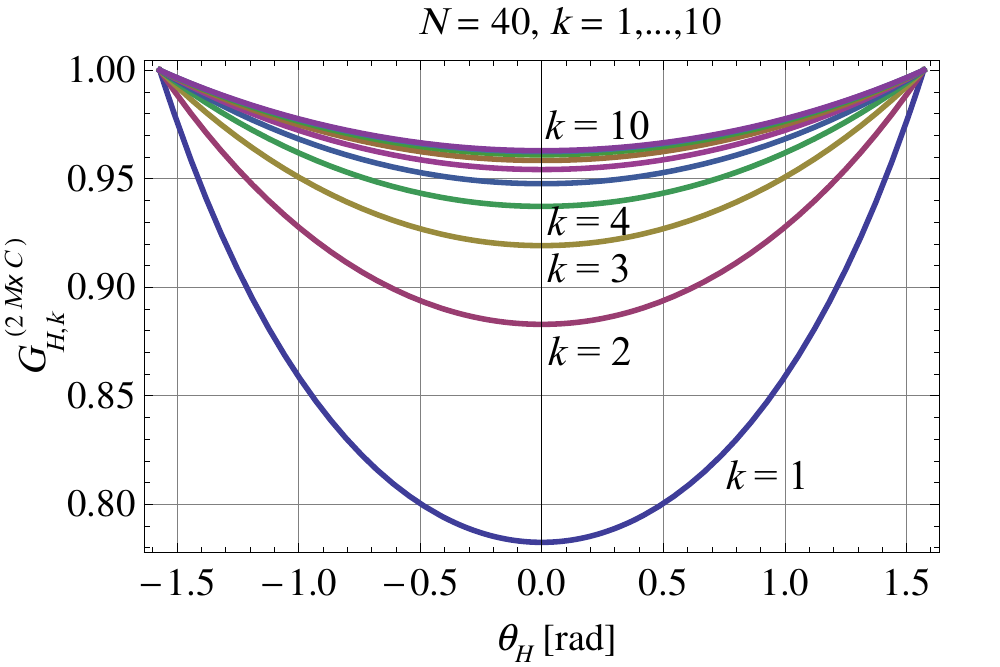}
    \caption{  Hall-geometry factor $G_{H,k}^{(2M\times C)}$ for a strictly regular Hall-plate with $40$ contacts ($M=20$), versus Hall-angle $\theta_H$ in the circuit of Fig. \ref{fig:single-input-current-multiple-output-voltages}. }
   \label{fig:GHk_2MxC_vs_thetaH}
\end{figure}

\noindent Fig. \ref{fig:GHk_2MxC_vs_thetaH} plots (\ref{eq:RMFoCD176}) for a Hall-plate with $40$ contacts versus Hall-angles $\theta_H\in[-\pi/2,\pi/2]$. We are interested in the Hall output signal between the two contacts, which are neighbors of the ground contact, $k\to 1$, for weak magnetic field and for a Hall-plate with many contacts. With (\ref{eq:RMFoCD171}) and (\ref{eq:RMFoCD130b}) this gives  
\begin{equation}\label{eq:RMFoCD177}\begin{split}
 \lim_{\theta_H\to 0}\lim_{M\to\infty} G_{H,1}^{(2M\times C)} & = \frac{4}{M} \sum_{m=1,3,5\ldots}^{M-1} \cos\left(\pi\frac{m}{2M}\right) \frac{\mathrm{arg}\{{^i}\!\gamma_m\}}{\theta_H} \\
& = \frac{8}{\pi^2}-\frac{8}{\pi}\sum_{k=0}^\infty \frac{k a_k}{4k^2-1} = 0.78341 ,
\end{split}\end{equation}
whereby we used $a_k$ from (\ref{eq:RMFoCD130c}) for $k=1,\ldots 20$ and checked the result against the exact formula (\ref{eq:RMFoCD130}) with $N=20000$. This limit in (\ref{eq:RMFoCD177}) is interesting, because the contact spacings become infinitely small, which pulls the Hall-geometry factor towards $0$. On the other hand the contacts also become infinitely small, which pulls the Hall-geometry factor towards $1$. The net result of both concurrent effects is a Hall-geometry factor of $0.78$. This agrees with Table 1 and Figure 7 in Ref.~\cite{Ausserlechner2020a}, both of which were obtained by numerical methods \cite{Homentcovschi2019}. The practical consequence of (\ref{eq:RMFoCD177}) is, that in Fig. \ref{fig:GHk_2MxC_vs_thetaH} there is no curve notably below the $k=1$ curve, even if $N$ gets larger than $40$.

For Hall-plates with an even number of contacts the input resistance $V_\mathrm{supply}/I_\mathrm{supply}$ at zero magnetic field in Fig. \ref{fig:single-input-current-multiple-output-voltages} is identical to the entry $R_{M,M}$ of the resistance matrix (with $M=N/2$). We insert (\ref{eq:RMFoCD55}) into (\ref{eq:RMFoCD45}) and after several manipulations it gives 
\begin{equation}\label{eq:RMFoCD88}
\frac{R_{M,M}}{R_\mathrm{sheet}} = \frac{1}{M}\sum_{k=1}^{M} \csc\left(\pi\frac{2k-1}{2M}\right) \text{ for }\theta_H=0 ,
\end{equation}
with $\csc (x) = 1/\sin (x)$. For large numbers of contacts one can show (again after several manipulations) 
\begin{equation}\label{eq:RMFoCD89}\begin{split}
\frac{R_{M,M}}{R_\mathrm{sheet}} & = \frac{2}{\pi}\left[ \gamma_{EM} + \ln\left(\frac{4}{\pi}\right) + \ln (N) \right] +\mathcal{O}\left(\frac{1}{N}\right) \\
& = 0.521252 + 0.63662 \ln (N) \text{ for }\theta_H=0 ,
\end{split}\end{equation}
with the Euler-Mascheroni constant $\gamma_{EM} \approx 0.577216$. Equation (\ref{eq:RMFoCD88}) is consistent with the values reported in Table 2 and Figure 8 in Ref.~\cite{Ausserlechner2020a}. The relative accuracy of (\ref{eq:RMFoCD89}) is $0.0074$ for $N=4$ and it improves for larger $N$. In Ref.~\cite{Ausserlechner2020a} it was conjectured that $R_{M,M}$ goes up logarithmically with the number of contacts, and in (\ref{eq:RMFoCD89}) this is strictly proven.

\section{The maximum noise efficiency of strictly regular Hall-plates}
\label{sec:noise-efficiency}

Here we discuss what is the maximum signal-to-thermal-noise ratio (SNR) per invested power of stricly regular Hall-plates. This question was investigated numerically for weak magnetic field in Refs.~\cite{Ausserlechner2020a} and \cite{Ausserlechner2020hybrid}, and in fact it kindled my interest in eigenvalues of Hall-plates. With the findings in this work we can consolidate the theory in Ref.~\cite{Ausserlechner2020hybrid} and give rigid limits for $N\to\infty$ for arbitrary magnetic field. 

\begin{figure}
  \centering
                \includegraphics[width=0.85\textwidth]{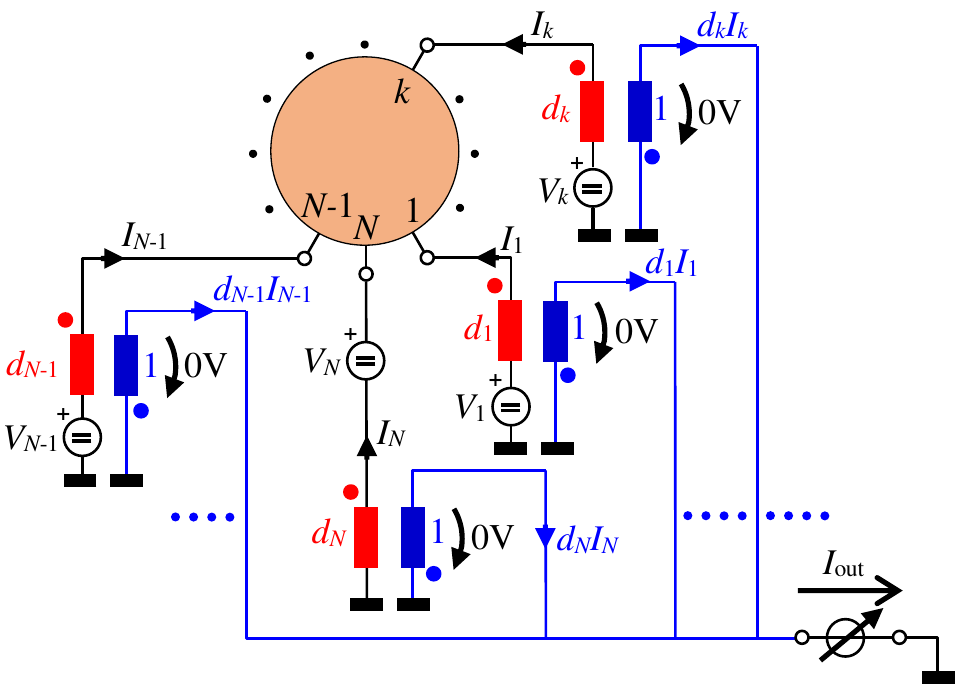}
    \caption{  Hall-plate with $N$ contacts in a conceptual circuit for current-mode operation. $N$ voltage sources $V_k$ at the contacts supply the Hall-plate with electric energy. $N$ passive noise-less ideal transformers tap the currents through the contacts. Each transformer has a dedicated turns ratio $d_k : 1$. The secondary sides of all transformers are connected in parallel to add up all currents. This sum is measured by an ideal amperemeter. Zero voltage across the ideal amperemeter forces zero voltage across primary and seconary sides of all transformers. (Figure adapted from Ref.~\cite{Ausserlechner2020hybrid}.)  }
   \label{fig:current-mode}
\end{figure}

In Ref.~\cite{Ausserlechner2020hybrid} we showed that in the limit of weak magnetic fields it is irrelevant if we apply voltages or currents or both of them to the contacts of a Hall-plate -- in all these cases we can achieve maximum SNR, if we choose appropriate patterns of voltages and currents. Therefore we can limit the present discussion to the case where supply voltages are connected to the contacts of a Hall-plate, and the currents in response to these voltages are sampled by amperemeters. This so-called \emph{current-mode operation} is shown in Fig. \ref{fig:current-mode}, where ideal transformers couple out the currents though the Hall-plate and collect them through a single amperemeter. Note that ideal transformers couple voltages and currents between their primary and secondary sides at arbitrary frequency -- even at 0 Hz. In practice one would use a different circuit instead of the ideal transformers (with additional minor losses in SNR and power), but for an investigation of the theoretical limits of a Hall-sensor circuit we prefer the loss-less and noise-less ideal transformers. Moreover, we assume \emph{two identical} sensor circuits, each one with its own Hall-plate, one exposed to positive magnetic field rendering the output current $I_\mathrm{out}(\theta_H)$, and the other one exposed to negative magnetic field rendering the output current $I_\mathrm{out}(-\theta_H)$. Finally, both are subtracted to give $I_\mathrm{diff}$. This complete circuit is called the \emph{difference-field circuit}, which we want to optimize with respect to signal over noise and power.

\begin{figure}
  \centering
                \includegraphics[width=0.73\textwidth]{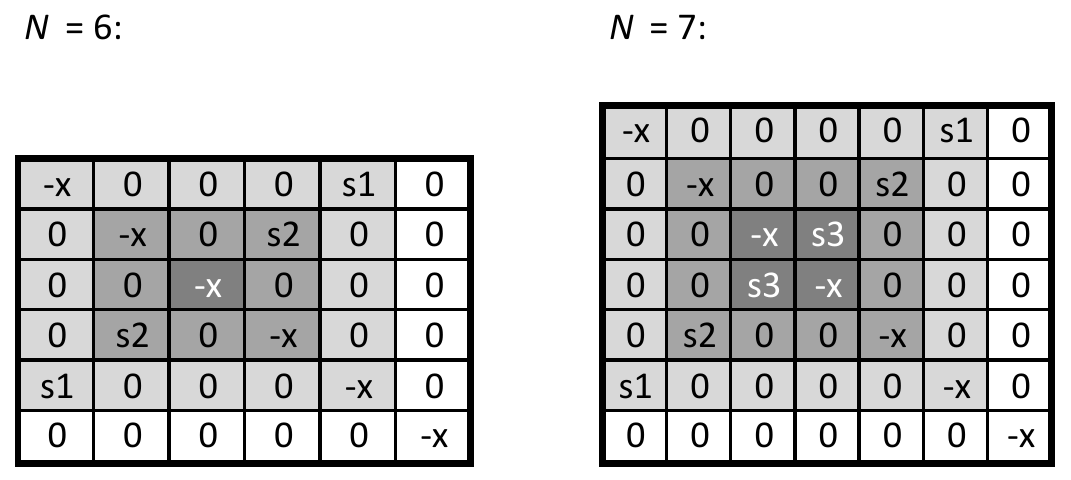}
    \caption{  Patterns of the matrix $(\bm{S}-x*\bm{1})$ to compute the eigenvalues of $\bm{S}$. The determinant of the $k$-th light gray-shaded square is equal to $(x^2-s_k^2)$ times the determinant of the $k+1$-th darker gray-shaded square. It finally holds $\mathrm{det}(\bm{S}-x*\bm{1}) = -x(x^2-s_1^2)(x^2-s_2^2)\cdots $. }
   \label{fig:eigenvalues_of_S}
\end{figure}

The output of the difference-field circuit can be written like 
\begin{equation}\label{eq:noise-efficiency1}
I_\mathrm{diff}=I_\mathrm{out}(\theta_H)-I_\mathrm{out}(-\theta_H) = \bm{d}^T\left( \bm{{^i}\!G} - \bm{{^i}\!G}^T \right) \bm{V}_\mathrm{supply} ,
\end{equation}
whereby we used the RMFR-principle\cite{Sample1987,Cornils2008}, $\bm{{^i}\!G}^T = \bm{{^i}\!G}(-\theta_H)$. The turns ratios of the ideal transformers are collected in the coefficient vector $\bm{d}$. Thus, the transformers not only couple out the currents, but they also implement a weighted sum of all currents (linear combination), whereby the first task of the optimization is to find a coefficient vector $\bm{d}$ that maximizes SNR over power. The power dissipated by the difference-field circuit is 
\begin{equation}\label{eq:noise-efficiency2}\begin{split}
P_\mathrm{diff} & =\bm{I}^T(\theta_H)\bm{V}_\mathrm{supply} + \bm{I}^T(-\theta_H)\bm{V}_\mathrm{supply} \\ 
& = \bm{V}_\mathrm{supply}^T \left(\bm{{^i}\!G} + \bm{{^i}\!G}^T \right) \bm{V}_\mathrm{supply} ,
\end{split}\end{equation}
whereby we used again the RMFR-principle\cite{Sample1987,Cornils2008}. In (\ref{eq:noise-efficiency1}) and (\ref{eq:noise-efficiency2}) we used the indefinite conductance matrix, because we want to use our findings from the preceding sections. Therefore, the vectors $\bm{d},\bm{V}_\mathrm{supply},\bm{I}$ have $N$ rows (this is in contrast to Ref.~\cite{Ausserlechner2020hybrid}). According to Nyquist and Johnson \cite{Nyquist,Johnson} the thermal noise current at the output of the difference-field circuit is determined by the output conductances 
\begin{equation}\label{eq:noise-efficiency3}\begin{split}
i_{\mathrm{diff},rms} = \sqrt{4 k_b T \Delta_f G_\mathrm{out}\left(\theta_H\right) + 4 k_b T \Delta_f G_\mathrm{out}\left(-\theta_H\right)} ,
\end{split}\end{equation}
with Boltzmann's constant $k_b$, the absolute temperature $T$, and the effective noise bandwidth $\Delta_f$. We neglect other noise sources, because they are partly absent in practice, and partly they can be eliminated by circuit techniques, e.g. contact commutation schemes or spinning current schemes. 
For the calculation of the output conductance we use Fig. \ref{fig:current-mode} and replace all voltage sources by shorts and we take the amperemeter away. Then we apply a test voltage $V_\mathrm{test}$ at the terminal, where the amperemeter was connected. The ideal transformers couple this voltage to the contacts of the Hall-plate, with $\bm{V} = -V_\mathrm{test}\bm{d}$. The Hall-plate exposed to positive magnetic field responds with currents $\bm{I} =\bm{{^i}\!G} \bm{V}$, which are again coupled out by the transformers and combined to $I_\mathrm{test} = -\bm{d}^T\bm{I}$, whereby $I_\mathrm{test}$ counts positive if it flows \emph{into} the Hall-plate. In total this gives 
\begin{equation}\label{eq:noise-efficiency4}
G_\mathrm{out} = \frac{I_\mathrm{test}}{V_\mathrm{test}} = \bm{d}^T \bm{{^i}\!G} \bm{d} .
\end{equation}
Combining (\ref{eq:noise-efficiency1}-\ref{eq:noise-efficiency4}) gives the ratio of signal to thermal noise  
\begin{equation}\label{eq:noise-efficiency5}
\mathrm{SNR} = \frac{\left|I_\mathrm{diff}\right|}{i_{\mathrm{diff},rms}} = \frac{\left|\bm{d}^T \left(\bm{{^i}\!G}-\bm{{^i}\!G}^T\right)\bm{V}_\mathrm{supply}\right|}{\sqrt{4 k_b T \Delta_f} \sqrt{ \bm{d}^T \left(\bm{{^i}\!G}+\bm{{^i}\!G}^T\right)\bm{d}}} .
\end{equation}
Obviously, the signal grows with larger Hall-angle and power, while the noise grows with higher temperature and larger system bandwidth. Therefore we normalize the SNR to get rid of all these obvious dependencies. 
\begin{equation}\label{eq:noise-efficiency6}\begin{split}
\eta' & = \frac{\mathrm{SNR}}{\left|\tan(\theta_H)\right|}\;\sqrt{\frac{4 k_b T \Delta_f}{P_\mathrm{diff}}} \\ 
& = \frac{1}{\sqrt{\bm{d}^T \,\bm{{^i}\!G}_\mathrm{ev} \bm{d}}} \frac{\left| \bm{d}^T \,\bm{{^i}\!G}_\mathrm{odd} \bm{V}_\mathrm{supply}\right|}{\sqrt{\bm{V}_\mathrm{supply}^T \bm{{^i}\!G}_\mathrm{ev} \bm{V}_\mathrm{supply}}} \frac{1}{\left|\tan(\theta_H)\right|} .
\end{split}\end{equation}
We call $\eta'$ the \emph{noise efficiency}. It is a figure of merit for the noise performance of a Hall-plate, which does \emph{not} depend on material properties like Hall-mobility or sheet resistance -- instead it depends on the layout (shape of the Hall-plate, number of contacts, sizes of contacts and their spacings), on the pattern of the supply voltages $\bm{V}_\mathrm{supply}$, and on the pattern of the coefficients $\bm{d}$. Note that a common scaling factor on all voltages or on all coefficients does not change $\eta'$. In principle, $\eta'$ is a function of the Hall-angle, but in practice we are interested in its limit for weak magnetic field, $\eta'_0$. With (\ref{eq:RMFoCD70}) we can write 
\begin{equation}\label{eq:noise-efficiency7}\begin{split}
& \bm{V}_\mathrm{supply}^T \,\bm{{^i}\!G}_\mathrm{ev} \bm{V}_\mathrm{supply} = \bm{V}_\mathrm{supply}^T \bm{Q}\,\bm{{^i}\!\Gamma_\mathrm{ev}}\bm{Q}^C \bm{V}_\mathrm{supply} \\ 
&\qquad =  \bm{V}_\mathrm{supply}^T \bm{Q}\,\bm{{^i}\!\Gamma_\mathrm{ev}}\bm{Q}^C \left(\bm{Q}^C\bm{Q}\right) \bm{V}_\mathrm{supply} \\ 
&\qquad =  \bm{V}_\mathrm{supply}^T \bm{Q}\bm{S}\bm{Q} \bm{V}_\mathrm{supply} \\ 
&\qquad = \left(\bm{S}^{1/2}\bm{Q}\bm{V}_\mathrm{supply}\right)^T  \left(\bm{S}^{1/2}\bm{Q}\bm{V}_\mathrm{supply}\right) , 
\end{split}\end{equation}
with the abbreviation 
\begin{equation}\label{eq:noise-efficiency7b}\begin{split}
\bm{S} = \bm{{^i}\!\Gamma_\mathrm{ev}}\bm{Q}^2 = \left(\begin{array}{ccccc} 0 & 0 & \reflectbox{$\ddots$} & {^i}\gamma_{\mathrm{ev},1} & 0 \\ \vdots & \reflectbox{$\ddots$} & \reflectbox{$\ddots$} & \reflectbox{$\ddots$} & \vdots \\ 0 & {^i}\gamma_{\mathrm{ev},N-2} & \reflectbox{$\ddots$} & 0 & 0 \\ {^i}\gamma_{\mathrm{ev},N-1} & 0 & \reflectbox{$\ddots$} & 0 & 0 \\ 0 & 0 & \cdots & 0 & 0  \end{array} \right) . 
\end{split}\end{equation}
\noindent In (\ref{eq:noise-efficiency7b}) we used (\ref{eq:RMFoCD67}). $\bm{S}$ is even because of (\ref{eq:RMFoCD72}). 
In (\ref{eq:noise-efficiency7}) we used the fact that $\bm{{^i}\!\Gamma_\mathrm{ev}}, \bm{Q},\bm{S}$ are even matrices. The eigenvalues of $\bm{S}$ are $0,\pm{^i}\gamma_{\mathrm{ev},1},\pm{^i}\gamma_{\mathrm{ev},2},\ldots$ as explained in Fig. \ref{fig:eigenvalues_of_S}. With the eigenvalues we can readily compute the eigenvectors. The result gives 
\begin{equation}\label{eq:noise-efficiency7c}\begin{split}
\bm{S} & = \bm{E} \;\mathrm{diag}({^i}\gamma_{\mathrm{ev},1},{^i}\gamma_{\mathrm{ev},2},\ldots,-{^i}\gamma_{\mathrm{ev},2},-{^i}\gamma_{\mathrm{ev},1},0) \;\bm{E}^{-1} , \\ 
& (\bm{E})_{k,\ell} = \delta_{k,\ell} + \mathrm{sign}(k-\ell) \delta_{k+\ell,N} , \\
& \bm{E} = \left(\begin{array}{ccccc}1&0&0&-1&0\\0&1&-1&0&0\\0&1&1&0&0\\1&0&0&1&0\\0&0&0&0&1\end{array}\right) \text{ for }N=5 , \\
& \bm{E} = \left(\begin{array}{cccccc}1&0&0&0&-1&0\\0&1&0&-1&0&0\\0&0&1&0&0&0\\0&1&0&1&0&0\\1&0&0&0&1&0\\0&0&0&0&0&1\end{array}\right) \text{ for }N=6 , 
\end{split}\end{equation}
with $\mathrm{sign}(x)$ equal to $-1,0,1$ depending on whether $x$ is negative, zero, or positive. The columns of $\bm{E}$ are the orthogonal eigenvectors, with $\mathrm{det}(\bm{E})=2^{\lfloor (N-1)/2\rfloor}$. %numerisch ueberprueft
This can be seen by adding the $(N-k)$-th row of $\bm{E}$ to its $k$-th row, for $1\le k\le\lfloor N/2\rfloor$ -- this procedure results in a lower triangular matrix, whose determinant is the product of entries on the main diagonal. Therefore, $\bm{E}$ is regular and its inverse exists. 
\begin{equation}\label{eq:noise-efficiency7d}\begin{split}
\bm{E}^{-1} & = \frac{1}{2} \left(\begin{array}{ccccc}1&0&0&1&0\\0&1&1&0&0\\0&-1&1&0&0\\-1&0&0&1&0\\0&0&0&0&2\end{array}\right) \text{ for }N=5 , \\
\bm{E}^{-1} & = \frac{1}{2} \left(\begin{array}{cccccc}1&0&0&0&1&0\\0&1&0&1&0&0\\0&0&2&0&0&0\\0&-1&0&1&0&0\\-1&0&0&0&1&0\\0&0&0&0&0&2\end{array}\right) \text{ for }N=6 . 
\end{split}\end{equation}
In (\ref{eq:noise-efficiency7c}) the main diagonal of the diagonal matrix has its first $\lfloor N/2\rfloor$ entries positive, whereas the entries with indices $\lfloor N/2\rfloor+1\ldots N-1$ are negative. Thus, $\bm{S}$ is not semi-definite. Nevertheless, its square-root is straightforward, 
\begin{equation}\label{eq:noise-efficiency7e}\begin{split}
\bm{S}^{1/2} = \bm{E} \;\mathrm{diag}\left(\sqrt{{^i}\gamma_{\mathrm{ev},1}},\sqrt{{^i}\gamma_{\mathrm{ev},2}},\ldots,\mathbbm{i}\sqrt{{^i}\gamma_{\mathrm{ev},2}},\mathbbm{i}\sqrt{{^i}\gamma_{\mathrm{ev},1}},0\right) \;\bm{E}^{-1} . 
\end{split}\end{equation}
In (\ref{eq:noise-efficiency7e}) we arbitrarily chose all positive square-roots. Here, the upper half of the diagonal matrix comprises real numbers, while the lower half comprises imaginary numbers. We introduce two vectors 
\begin{equation}\label{eq:noise-efficiency8}\begin{split}
\bm{v}_1 = \bm{S}^{1/2}\bm{Q}\bm{V}_\mathrm{supply} \quad\text{and}\quad \bm{v}_2 = \bm{S}^{1/2}\bm{Q}\bm{d} . 
\end{split}\end{equation}
Note that $\bm{E}^{-1}\bm{Q}$ has real values in its top $\lfloor N/2\rfloor$ rows and its $N$-th row whereas all other rows have only imaginary or zero entries. If we use this in (\ref{eq:noise-efficiency7e}), it follows that $\bm{S}^{1/2}\bm{Q}$ is real valued. Therefore $\bm{v}_1,\bm{v}_2$ in (\ref{eq:noise-efficiency8}) are real valued, because $\bm{V}_\mathrm{supply},\bm{d}$ are real valued. Consequently, the square-roots in  (\ref{eq:noise-efficiency6}) are the norms of $\bm{v}_1,\bm{v}_2$ according to (\ref{eq:noise-efficiency7}) and (\ref{eq:noise-efficiency8}), i.e., the arguments of the square-roots are indeed positive. Unfortunately, $\bm{S}$ is singular, because $\bm{{^i}\!\Gamma_\mathrm{ev}}$ is singular. Nevertheless, we can invert both equations in (\ref{eq:noise-efficiency8}) by the pseudo-inverse (Moore-Penrose inverse, denoted by the superscript ${^+}$), 
\begin{equation}\label{eq:noise-efficiency9}\begin{split}
\bm{V}_\mathrm{supply} = \bm{Q}^C \left(\bm{S}^{1/2}\right)^+\bm{v}_1 ,\quad \bm{d} = \bm{Q}^C\left(\bm{S}^{1/2}\right)^+\bm{v}_2 , 
\end{split}\end{equation}
whereby the pseudo-inverse is readily given by replacing all non-vanishing entries $x$ by $1/x$ in the diagonal matrix of (\ref{eq:noise-efficiency7e}), 
\begin{equation}\label{eq:noise-efficiency10}\begin{split}
& \left(\bm{S}^{1/2}\right)^+ = \bm{E} \;\mathrm{diag}\left(\left({^i}\gamma_{\mathrm{ev},1}\right)^{-1/2},\left({^i}\gamma_{\mathrm{ev},2}\right)^{-1/2},\ldots \right. \\
& \qquad\qquad \left. \ldots,-\mathbbm{i}\left({^i}\gamma_{\mathrm{ev},2}\right)^{-1/2},-\mathbbm{i}\left({^i}\gamma_{\mathrm{ev},1}\right)^{-1/2},0\right) \;\bm{E}^{-1} \\ 
& = \frac{1}{2} \left(\begin{array}{cccccc}\frac{1-\mathbbm{i}}{\sqrt{{^i}\gamma_{\mathrm{ev},1}}}&0&\cdots&0&\frac{1+\mathbbm{i}}{\sqrt{{^i}\gamma_{\mathrm{ev},1}}}&0\\0&\frac{1-\mathbbm{i}}{\sqrt{{^i}\gamma_{\mathrm{ev},2}}}&\ddots&\frac{1+\mathbbm{i}}{\sqrt{{^i}\gamma_{\mathrm{ev},2}}}&0&0\\ \vdots& \ddots &\ddots&\ddots&\vdots&\vdots\\0&\frac{1+\mathbbm{i}}{\sqrt{{^i}\gamma_{\mathrm{ev},2}}}&\reflectbox{$\ddots$}&\frac{1-\mathbbm{i}}{\sqrt{{^i}\gamma_{\mathrm{ev},2}}}&0&0\\\frac{1+\mathbbm{i}}{\sqrt{{^i}\gamma_{\mathrm{ev},1}}}&0&\cdots&0&\frac{1-\mathbbm{i}}{\sqrt{{^i}\gamma_{\mathrm{ev},1}}}&0\\0&0&\cdots&0&0&0\end{array}\right) . 
\end{split}\end{equation}
For even $N$ the two non-vanishing diagonals in $\left(\bm{S}^{1/2}\right)^+$ in (\ref{eq:noise-efficiency10}) meet in the entry with column and row indices $N/2$, and then this entry equals $1/\sqrt{{^i}\gamma_{\mathrm{ev},N/2}}$. 
In fact (\ref{eq:noise-efficiency9}) gives only one solution for $\bm{V}_\mathrm{supply}, \bm{d}$ out of infinitely many solutions of (\ref{eq:noise-efficiency8}). For voltages, it is well-known that the choice of the ground node does neither change the signal swing, nor the noise or the power of an electric system. The same applies to the coefficient vector $\bm{d}$: the SNR of the Hall-plate does not change, if we add a constant value to all entries of $\bm{d}$ (this was also explained in detail in Ref.~\cite{Ausserlechner2020hybrid}). Inserting (\ref{eq:noise-efficiency9}) into (\ref{eq:noise-efficiency6}) gives 
\begin{equation}\label{eq:noise-efficiency11}\begin{split}
\eta' & = \frac{\left|\bm{v}_2^T \bm{M} \bm{v}_1\right|}{\sqrt{\bm{v}_1^T \bm{v}_1} \sqrt{\bm{v}_2^T \bm{v}_2}} \\ 
& \text{with }\bm{M} = \left(\bm{S}^{1/2}\right)^+ \frac{\bm{{^i}\!\Gamma_\mathrm{odd}}}{\tan(\theta_H)} \bm{Q}^2 \left(\bm{S}^{1/2}\right)^+ . 
\end{split}\end{equation}
$\bm{{^i}\!\Gamma_\mathrm{odd}}\bm{Q}^2$ is analogous to $\bm{{^i}\!\Gamma_\mathrm{ev}}\bm{Q}^2$ if we replace ${^i}\gamma_{\mathrm{ev},k}$ by ${^i}\gamma_{\mathrm{odd},k}$ in (\ref{eq:noise-efficiency7b}). Multiplying out the matrices in (\ref{eq:noise-efficiency11}) gives the odd-symmetric real-valued matrix $\bm{M}$, 
\begin{equation}\label{eq:noise-efficiency11a}\begin{split}
\bm{M} = \frac{\mathbbm{i}}{\tan(\theta_H)} \left(\begin{array}{ccccc} 0 & 0 & \reflectbox{$\ddots$} & \frac{-{^i}\gamma_{\mathrm{odd},1}}{{^i}\gamma_{\mathrm{ev},1}} & 0 \\ \vdots & \reflectbox{$\ddots$} & \reflectbox{$\ddots$} & \reflectbox{$\ddots$} & \vdots \\ 0 & \frac{{^i}\gamma_{\mathrm{odd},2}}{{^i}\gamma_{\mathrm{ev},2}} & \reflectbox{$\ddots$} & 0 & 0 \\ \frac{{^i}\gamma_{\mathrm{odd},1}}{{^i}\gamma_{\mathrm{ev},1}} & 0 & \reflectbox{$\ddots$} & 0 & 0 \\ 0 & 0 & \cdots & 0 & 0  \end{array} \right) . 
\end{split}\end{equation}
We apply Schwartz' inequality 
\begin{equation}\label{eq:noise-efficiency12}\begin{split}
\left|\bm{v}_2^T \bm{M} \bm{v}_1\right| = \left|\left(\bm{M}^T \bm{v}_2\right)^T \bm{v}_1\right| \le \sqrt{\left(\bm{M}^T \bm{v}_2\right)^T \left(\bm{M}^T \bm{v}_2\right)} \sqrt{\bm{v}_1^T \bm{v}_1} , 
\end{split}\end{equation}
whereby the maximum value is obtained for colinear vectors, 
\begin{equation}\label{eq:noise-efficiency12b}
\bm{v}_1 = s \bm{M}^T \bm{v}_2 , 
\end{equation}
with some real number $s$. Inserting this into (\ref{eq:noise-efficiency11}) gives
\begin{equation}\label{eq:noise-efficiency13}\begin{split}
\left(\eta'_\mathrm{max}\right)^2 = \frac{\bm{v}_2^T \bm{M}\bm{M}^T \bm{v}_2}{\bm{v}_2^T \bm{v}_2} . 
\end{split}\end{equation}
The right hand side of (\ref{eq:noise-efficiency13}) is a Rayleigh quotient, whose maximum equals the largest magnitude of the eigenvalues of the symmetric matrix $\bm{M}\bm{M}^T$, and it is obtained, if $\bm{v}_2$ equals the corresponding eigenvector \cite{Rayleigh-Quotient,Horn}. From (\ref{eq:noise-efficiency11a}) we get by simple matrix multiplication 
\begin{equation}\label{eq:noise-efficiency13b}\begin{split}
\bm{M}\bm{M}^T = \left(\!\frac{\mathbbm{i}}{\tan(\theta_H)}\!\right)^{\!\!2} \left(\begin{array}{ccccc} \left(\frac{{^i}\gamma_{\mathrm{odd},1}}{{^i}\gamma_{\mathrm{ev},1}}\right)^2 & 0 & \ddots & 0 & 0 \\  0 & \left(\frac{{^i}\gamma_{\mathrm{odd},2}}{{^i}\gamma_{\mathrm{ev},2}}\right)^2 & \cdots & 0 & 0 \\ \vdots & \ddots & \ddots & \ddots & \vdots \\ \vdots & \ddots & \ddots & \left(\frac{{^i}\gamma_{\mathrm{odd},1}}{{^i}\gamma_{\mathrm{ev},1}}\right)^2 & 0 \\ 0 & 0 & \ddots & 0 & 0 \end{array} \right) . 
\end{split}\end{equation}
Luckily, this is a diagonal matrix, with the eigenvalues on the main diagonal. The largest eigenvalue is the first and the $N-1$-th entry on the diagonal (see Fig. \ref{fig:argigamma-survey}). This finally gives the maximum noise efficiency of strictly regular Hall-plates 
\begin{equation}\label{eq:noise-efficiency14}\begin{split}
\eta'_\mathrm{max} = \left|\frac{1}{\tan(\theta_H)}\,\frac{{^i}\gamma_{\mathrm{odd},1}}{{^i}\gamma_{\mathrm{ev},1}}\right| = \frac{\tan\left( \mathrm{arg}\{{^i}\!\gamma_1\} \right)}{\tan(\theta_H)} . 
\end{split}\end{equation}
With (\ref{eq:RMFoCD131}) we get the ultimate limit of the noise-efficiency that we can draw out of a Hall-plate at weak magnetic field, 
\begin{equation}\label{eq:noise-efficiency15}
1 > \eta'_{0\mathrm{max}}> 1-\frac{2}{N}, \quad\text{with}\quad \lim_{N\to\infty} \eta'_{0\mathrm{max}} = 1 . 
\end{equation}
In Ref.~\cite{Ausserlechner2020hybrid} we obtained $\eta'_0 = 0.94239$ for strictly regular Hall-plates with $N=40$ contacts, while (\ref{eq:noise-efficiency15}) gives $\eta'_0 \approx 0.95$ and the exact formulas (\ref{eq:noise-efficiency14}), (\ref{eq:RMFoCD130}) give $\eta'_0 = 0.9423901576$. In Ref.~\cite{Ausserlechner2020hybrid} we could not study Hall-plates with more than 40 contacts due to numerical problems. Now we are able to show that the exact limit is $\eta'_0\to 1$, although the convergence is slow. Please note that conventional Hall-plates with four contacts have $\eta'_0=\sqrt{2}/3=0.471$ (insert (\ref{eq:RMFoCD162}), (\ref{eq:RMFoCD163}) into (\ref{eq:noise-efficiency14}), see also Refs.~\cite{AusserlechnerSNR2017,Ausserlechner2020a}). Thus, optimum Hall-plates with infinitely many contacts have $112\%$ larger noise efficiency at weak magnetic field than traditional Hall-plates with four contacts and strictly regular symmetry! 

\begin{figure}
  \centering
                \includegraphics[width=0.85\textwidth]{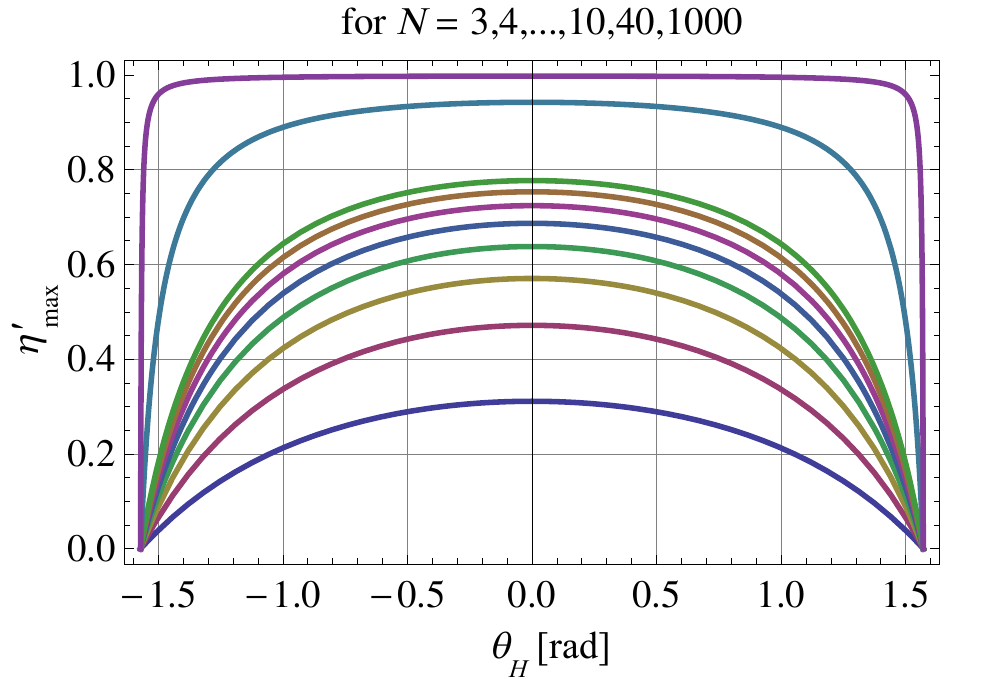}
    \caption{  Maximum possible noise efficiency $\eta'_\mathrm{max}$ for strictly regular Hall-plates with $N=3,4,\ldots ,10,40,1000$ contacts versus Hall-angle $\theta_H$. The lowest curve is for $N=3$, the top curve is for $N=1000$. The curves are given by (\ref{eq:noise-efficiency14}) and (\ref{eq:RMFoCD128}). In order to achieve $\eta'_\mathrm{max}$ the circuit concept of Fig. \ref{fig:current-mode} needs to be implemented with optimized patterns of supply voltages and coefficients according to (\ref{eq:noise-efficiency21}). }
   \label{fig:noise-efficiency}
\end{figure}

Next we discuss the noise efficiency at moderate and large magnetic fields. Thereby, we have to keep in mind that the noise efficiency basically is the normalized signal, $I_\mathrm{diff}$, which gives the change in Hall-plate output current upon \emph{reversal} of the magnetic field -- it is \emph{not} the change in output caused by a small change in magnetic field superimposed on a large constant magnetic field! 
For large contact count the noise efficiency tends to remain high up to larger magnetic fields, until at very strong magnetic field, $\theta_H\to\pm\pi/2$, it follows $\eta'\to 0$ according to (\ref{eq:RMFoCD129}) and (\ref{eq:noise-efficiency14}). Fig. \ref{fig:noise-efficiency} shows the maximum noise efficiency versus Hall-angle for strictly regular Hall-plates for various numbers of contacts. Clearly, for small contact count the noise efficiency drops faster versus magnetic field than for large numbers of contacts. The meaning of the small noise efficiency at strong magnetic field is that the geometrical magneto-resistance gets so large that the thermal noise dominates over the magnetic field response of the Hall-plate. Hence, the current mode circuit of Fig. \ref{fig:current-mode} cannot resolve large magnetic fields.

What are the optimum patterns for $\bm{V}_\mathrm{supply},\bm{d}$ for maximum noise efficiency? For $\eta'=\eta'_\mathrm{max}$ the vector $\bm{v}_2$ must be the first eigenvector of $\bm{M}\bm{M}^T$, which simply means 
\begin{equation}\label{eq:noise-efficiency20}
\bm{v}_2^T = (1,0,\ldots ,0), \quad \bm{v}_1^T = (0,\ldots ,0,1,0) , 
\end{equation}
where we discarded irrelevant real-valued pre-factors. In (\ref{eq:noise-efficiency20}) we obtained $\bm{v}_1$ by inserting $\bm{v}_2$ into (\ref{eq:noise-efficiency12b}) and using (\ref{eq:noise-efficiency11a}).
Inserting (\ref{eq:noise-efficiency20}) into (\ref{eq:noise-efficiency9}) gives 
\begin{equation}\label{eq:noise-efficiency21}
\bm{V}_\mathrm{supply} \!=\!\! \left(\!\begin{array}{c}\!\cos\left(2\pi\frac{1}{N}\!-\!\frac{\pi}{4}\right)\!\\ \!\cos\left(2\pi\frac{2}{N}\!-\!\frac{\pi}{4}\right)\!\\ \vdots\\ \!\cos\left(2\pi\frac{N}{N}\!-\!\frac{\pi}{4}\right)\!\end{array}\!\!\right)\!, \bm{d} \!=\!\! \left(\!\begin{array}{c}\!\cos\left(2\pi\frac{1}{N}\!+\!\frac{\pi}{4}\right)\!\\ \!\cos\left(2\pi\frac{2}{N}\!+\!\frac{\pi}{4}\right)\!\\ \vdots\\ \!\cos\left(2\pi\frac{N}{N}\!+\!\frac{\pi}{4}\right)\!\end{array}\!\!\right) 
\end{equation}

% original file: noise-eficiency_of_strictly_regular_Halls.nb

\noindent Note that both $\bm{V}_\mathrm{supply}$ and $\bm{d}$ do not depend on the magnetic field. Thus, any system with specific patterns for $\bm{V}_\mathrm{supply},\bm{d}$, which is optimum for noise at weak magnetic field, has also optimum noise-efficiency at arbitrary magnetic field. Since the Hall-plate has regular symmetry, we can add an integer multiple of $2\pi/N$ to all arguments of the cosines in (\ref{eq:noise-efficiency21}). Thus, the optimum patterns of supply voltages and coefficients are sinusoidal, whereby the supply voltages are maximum and minimum at opposite contacts, say $C_{\lfloor N/2\rfloor}, C_N$. There the coefficients are near zero. Conversely the coefficients are maximum and minimum at opposite contacts $C_{\lfloor N/4\rfloor}, C_{\lfloor 3N/4\rfloor}$ where the supply voltages are in the middle between maximum and minimum supply voltages. Hence, the patterns of $\bm{V}_\mathrm{supply}$ and $\bm{d}$ are shifted against each other by a quarter of the perimeter. If we supply constant currents instead of constant voltages, we use $\bm{I}_\mathrm{supply} = \bm{{^i}G_{\mathrm{ev}0}}\bm{V}_\mathrm{supply}$ according to Ref.~\cite{Ausserlechner2020hybrid}. Then it is straightforward to show that this supply current has the same sinusoidal pattern as the supply voltage in (\ref{eq:noise-efficiency21}). These patterns match perfectly with the ones found by numerical methods \cite{Ausserlechner2020hybrid}. There, we could not explain why the patterns for the best noise efficiency were sinusoidal. However, the theory developed in this work shows that the sinusoidal shape for the optimum patterns are a consequence of the fact that the indefinite conductance matrix is circulant (the cosine-dependence comes from the $\bm{Q}$-matrix in (\ref{eq:RMFoCD52})).

\begin{figure}
  \centering
                \includegraphics[width=0.85\textwidth]{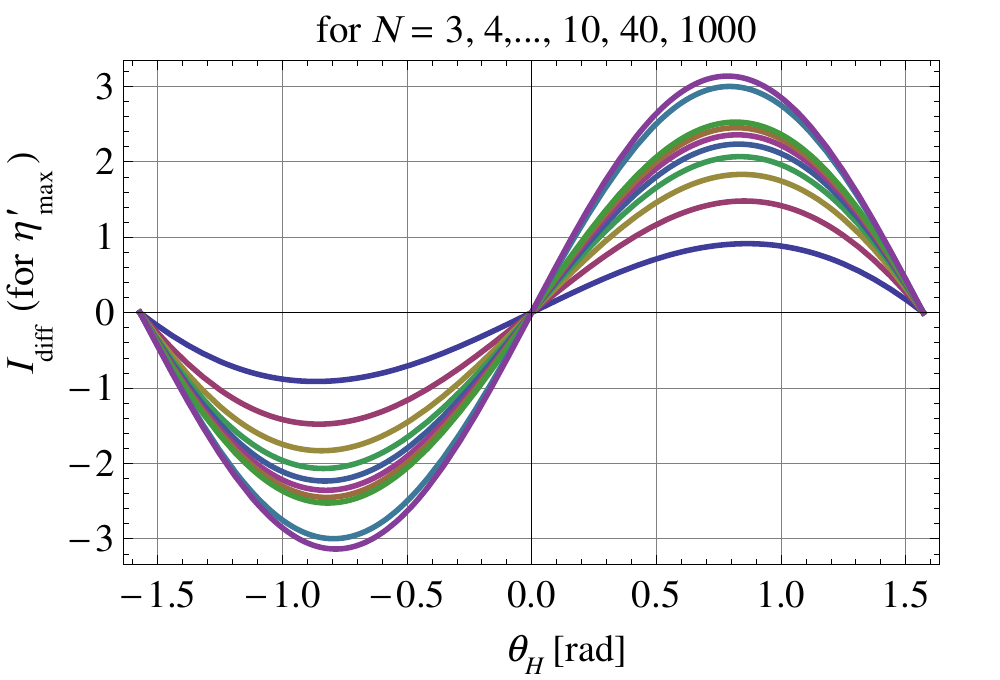}
    \caption{  Output currents $I_\mathrm{diff}=I_\mathrm{out}(\theta_H)-I_\mathrm{out}(-\theta_H)$ versus Hall-angle $\theta_H$ for current mode operation with constant supply voltages according to Fig.\ref{fig:current-mode}. The Hall-plates are strictly regular with $N=3,4,\ldots ,10,40,1000$ contacts, whereby the curve with largest amplitude has $N=1000$,, and fewer contacts mean smaller amplitude. We assume optimized patterns of supply voltages, (\ref{eq:noise-efficiency21}), for maximum noise efficiency $\eta'_\mathrm{max}$. The output currents were computed with (\ref{eq:noise-efficiency25}). They are non-monotonic versus Hall-angle. }
   \label{fig:Idiff}
\end{figure}

\begin{figure}
  \centering
                \includegraphics[width=0.85\textwidth]{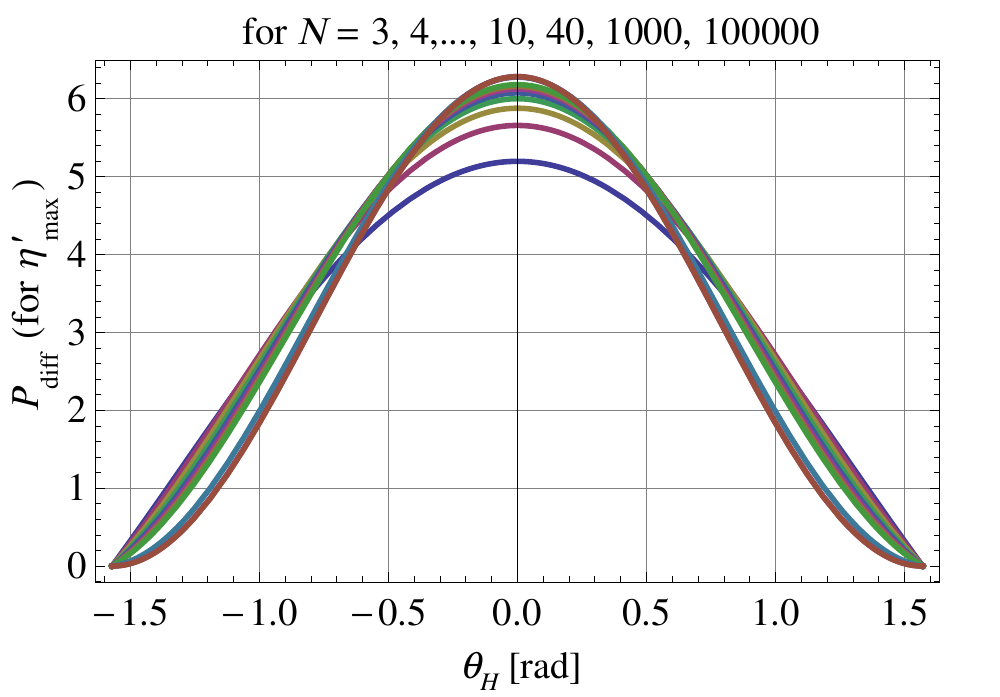}
    \caption{  Power $P_\mathrm{diff}$ of the Hall-plates from Fig. \ref{fig:Idiff} versus Hall-angle $\theta_H$ for current mode operation with constant supply voltages according to Fig. \ref{fig:current-mode}. The curve with largest amplitude has $N=100000$ contacts, and fewer contacts mean smaller amplitude. We assume optimized patterns for supply voltages for maximum noise efficiency $\eta'_\mathrm{max}$. The power was computed with (\ref{eq:noise-efficiency25}). In the limit $\theta_H\to\pm\pi/2$ all conductances of the Hall-plates vanish and therefore the plates dissipate no power. }
   \label{fig:Pdiff}
\end{figure}

Inserting the optimum patterns from (\ref{eq:noise-efficiency21}) into (\ref{eq:noise-efficiency1}) and (\ref{eq:noise-efficiency2}) gives the output current $I_\mathrm{diff}$, the dissipated power $P_\mathrm{diff}$, and the output conductance $G_\mathrm{out}$ of the difference-field circuit versus applied magnetic field, see Figs. \ref{fig:Idiff}, \ref{fig:Pdiff}, 
\begin{equation}\label{eq:noise-efficiency25}\begin{split}
& I_\mathrm{diff} = \frac{2}{R_\mathrm{sq}} N\sin\left(\frac{\pi}{N}\right) \sin\left(\mathrm{arg}\{{^i}\!\gamma_1\}\right) , \\ 
& P_\mathrm{diff} = N {^i}\!\gamma_{\mathrm{ev},1} = \frac{2}{R_\mathrm{sq}} N\sin\left(\frac{\pi}{N}\right) \cos\left(\mathrm{arg}\{{^i}\!\gamma_1\}\right) , \\
& G_\mathrm{out}(\theta_H)+G_\mathrm{out}(-\theta_H) = N {^i}\!\gamma_{\mathrm{ev},1} . 
\end{split}\end{equation}

% original file: noise-eficiency_of_strictly_regular_Halls.nb

\begin{figure}
  \centering
                \includegraphics[width=0.85\textwidth]{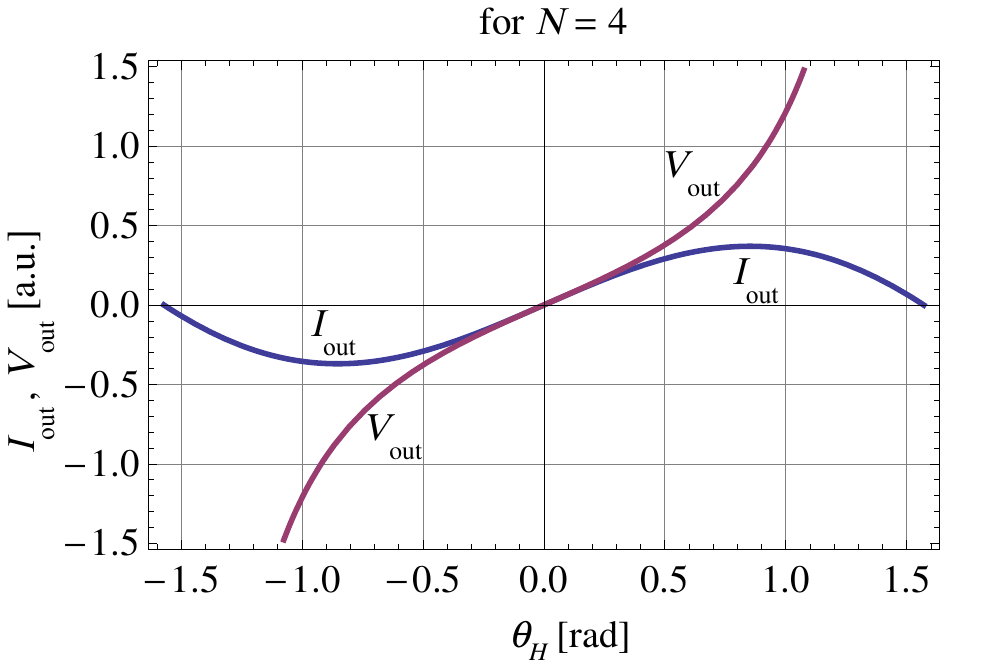}
    \caption{  Comparison of current mode operation versus voltage mode operation up to strong magnetic fields. The Hall-plate is strictly regular with $N=4$ contacts. In current mode operation according to Fig. \ref{fig:current-mode} a potential of $1$V is applied to $C_2,C_3$ and $-1$V is applied to $C_1,C_4$, and all currents are measured to compute $I_\mathrm{out}=(I_1+I_2-I_3-I_4)/4$. In an alternative current mode operation a voltage of $2$V is applied between $C_2-C_4$, and an amperemeter is connected between $C_3-C_1$ to measure $I_\mathrm{out}=I_1$ (For Hall-plates with four contacts this is easier to implement, and it has equivalent noise efficiency). In voltage mode operation a current of $1$A is injected into $C_2$ and extracted at $C_4$, and a voltmeter is connected between $C_3-C_1$ to measure $V_\mathrm{out}$. For both current mode operations the output current goes $\propto\sin\left(\mathrm{arg}\{{^i}\!\gamma_1\}\right)\cos(\theta_H)$, whereas in the voltage mode operation the output voltage goes $\propto\sin\left(\mathrm{arg}\{{^i}\!\gamma_1\}\right)/ \cos(\theta_H)$. Therefore the output signal in current mode is non-monotonic versus $\theta_H$ if $|\theta_H|$ exceeds $48.86$°, whereas the output signal in voltage mode is monotonic for all Hall-angles. }
   \label{fig:I-vs-V-mode4C}
\end{figure}

\noindent The Hall-output current $I_\mathrm{diff}$ is non-monotonic versus magnetic field -- for $|\theta_H|\approx > \pi/4$ it decreases with increasing magnetic field. Hence, for a magnetic field sensing application we have to limit the Hall-angle to be less than $\pm 45$°. This is a specific drawback of the current mode operation of Hall-plates, where constant voltages are supplied and currents are measured. Conversely, in voltage mode operation, constant supply currents are forced at the contacts and the potentials are tapped and added up in a linear combination to give $V_\mathrm{out}$. Then the output voltage $V_\mathrm{diff}=V_\mathrm{out}(\theta_H)-V_\mathrm{out}(-\theta_H)$ is monotonic versus Hall-angle for arbitrary Hall-angles. However, the voltage mode operation suffers from another drawback: It needs infinite voltages and infinite power to drive the constant currents through the Hall-plate at $\theta_H\to\pm\pi/2$ (which in practice means unacceptably high electric field). This is explained in Fig. \ref{fig:I-vs-V-mode4C} for a conventional Hall-plate with four contacts.

Note that a practical circuit can easily generate $I_\mathrm{diff}$ by subtracting $I_{\mathrm{out}1}-I_{\mathrm{out}2}$, whereby $I_{\mathrm{out}1}=I_{\mathrm{out}}(\theta_H)=\bm{d}^T\bm{{^i}\!G}\bm{V}_\mathrm{supply}$ like in (\ref{eq:noise-efficiency1}), whereas $I_{\mathrm{out}2}$ is obtained from a second Hall-plate connected identically like the first one, and then flipped upside down with its connections remaining,  $I_{\mathrm{out}2}=I_{\mathrm{out}}(-\theta_H)=\bm{d}^T\bm{{^i}\!G}(-\theta_H)\bm{V}_\mathrm{supply}$. Because of the symmetry of the regular Hall-plate we can also use the first Hall-plate instead of flipping a second one. Then all supply voltages $V_k$ and all ideal transformers with turns ratio $d_k$ have to be moved from contacts $C_k$ to $C_{N-k}$ for all $k\in\{1,2,\ldots N-1\}$. With the RMFR-principle this means $I_{\mathrm{out}2}=\bm{d}^T\bm{{^i}\!G}^T\bm{V}_\mathrm{supply}$. 
Finally, we can operate the first Hall-plate in a contact commutation scheme, where it is operated in a first clock cycle to deliver $I_{\mathrm{out}1}$, and in a second clock cycle the contacts are swapped to deliver $(-I_{\mathrm{out}2})$. If we send these data through a low-pass filter it gives $I_\mathrm{diff}$. In particular, it removes offset errors due to small asymmetries of the Hall-plate, $1/f$-noise in the frequency band between zero and half of the clock frequency, and planar Hall-voltages. Thus, the circuitry for the contact commutation scheme is a bit of work and effort, but it greatly improves the accuracy of the Hall-magnetic-field-sensor.

% original file: noise-eficiency_of_strictly_regular_Halls.nb

% original file: noise-eficiency_of_strictly_regular_Halls.nb

\section{Conclusion}
\label{sec:conclusion}
If Hall-plates are used as magnetic field sensors, the highest degree of symmetry is preferable, because it minimizes  thermal noise \cite{AusserlechnerSNR2017}. This calls for strictly regular Hall-plates, for which this work gives a very economical method of computation. Such a symmetric Hall-plate with $N$ contacts has only $N$ real numbers, which define the real and imaginary parts of the $N$ complex conjugate eigenvalues. The magnitudes of the eigenvalues are simple, (\ref{eq:RMFoCD69}), but the Hall-output signal is determined by the arguments of the eigenvalues in (\ref{eq:RMFoCD128}), which are given by integrals, that cannot be solved analytically. However, they are simple to compute numerically for arbitrary magnetic field. Especially for large Hall-angles this is an advantage of our formulae over other methods of computation for Hall-plates. With the eigenvalues, we know the complete conductance and resistance matrices as linear combinations thereof, i.e., we know all impedances and all Hall-geometry factors of the Hall-plate. Up to the author's knowledge there is no comparable method which gives closed analytical formulae for all electrical parameters of linear Hall-plates with arbitrary numbers of extended contacts and at arbitrary Hall-angles -- albeit it is limited to strictly regular symmetry. An extension of the proposed theory to regular symmetry is possible \cite{eigenvalues-regular}. 

New simple and accurate formulae for the Hall-geometry factors of strictly regular Hall-plates with three and four contacts were given as functions of the Hall-angle, (\ref{eq:RMFoCD154}), (\ref{eq:RMFoCD164}).
The complex eigenvalue of the indefinite conductance matrix with the largest argument determines the maximum noise efficiency of the Hall-plate. It can be achieved by a single period of a cosine-pattern of supply voltages along the boundary, and by a single period of a sine-pattern of weighting coefficients in the linear combination of output currents. For large numbers of contacts the noise efficiency tends to $1$ without exceeding it. This noise efficiency is up to $112$\% larger than the best noise efficiency one can draw out of traditional Hall-plates with four contacts. 

The eigenvalues of Hall-plates are one of the rare occasions in physics, where a complex quantity is not just a convenient mathematical tool for computation, but it bears real physical meaning. \\

\noindent This article is available in better formatting at DOI:10.13140/RG.2.2.28905.36968

\section*{Acknowledgments}
I am grateful to two anonymous reviewers for their guidance and for their scrupulous check of the original manuscript which I submitted to the Journal of Applied Physics, even though the journal rejected the manuscript as 'too mathematical' (compare with \cite{Wick1954} in the same journal). 

\appendix

%\section*{How to compute the sum in (\ref{eq:RMFoCD125})}
\section*{How to compute the sum in (58)}
\label{sec:sum}

We compute the sum 
\begin{equation}\label{eq:sum1}\begin{split}
\sum_{k=1}^N \frac{\exp\left(2\pi \mathbbm{i} k m / N\right)}{\sin(x/N\!+\!\pi k/N)\sin(x/N\!+\!\pi (k\!-\!j)/N)} ,
\end{split}\end{equation} 
for $N\ge 2$, $j\in\{1,2,\ldots,N-1\}$, and integer $m$.
First we replace $k=N$ by $k=0$, then we express the sines by complex exponentials, 
\begin{equation}\label{eq:sum2}\begin{split}
& \sum_{k=0}^{N-1} \frac{(2\mathbbm{i})^2 \exp\left(2\pi \mathbbm{i} k m / N\right)}{\exp\left(\mathbbm{i}(x/N\!+\!\pi k/N)\right) \exp\left(\mathbbm{i}(x/N\!+\!\pi (k\!-\!j)/N)\right)} \\
&\qquad * \left[1\!-\!\exp\left(-2\mathbbm{i}\frac{x\!+\!\pi k}{N}\right)\right]^{-1} \left[1\!-\!\exp\left(-2\mathbbm{i}\frac{x\!+\!\pi (k\!-\!j)}{N}\right)\right]^{-1} .
\end{split}\end{equation} 
$*$ means a scalar multiplication. We employ a double Taylor series expansion of the type $[1-z]^{-1}=1+z+z^2+z^3+\ldots$. This gives 
\begin{equation}\label{eq:sum3}\begin{split}
& \sum_{\ell=0}^\infty\sum_{q=0}^\infty -4\exp\left(-2\mathbbm{i}x\frac{1+\ell+q}{N}\right) \exp\left(\mathbbm{i}\pi j\frac{1+2q}{N}\right) \\
& \qquad\quad * \sum_{k=0}^{N-1} \exp\left(2\pi \mathbbm{i}k\frac{m-1-\ell-q}{N}\right) .
\end{split}\end{equation} 
\begin{figure}
  \centering
                \includegraphics[width=0.30\textwidth]{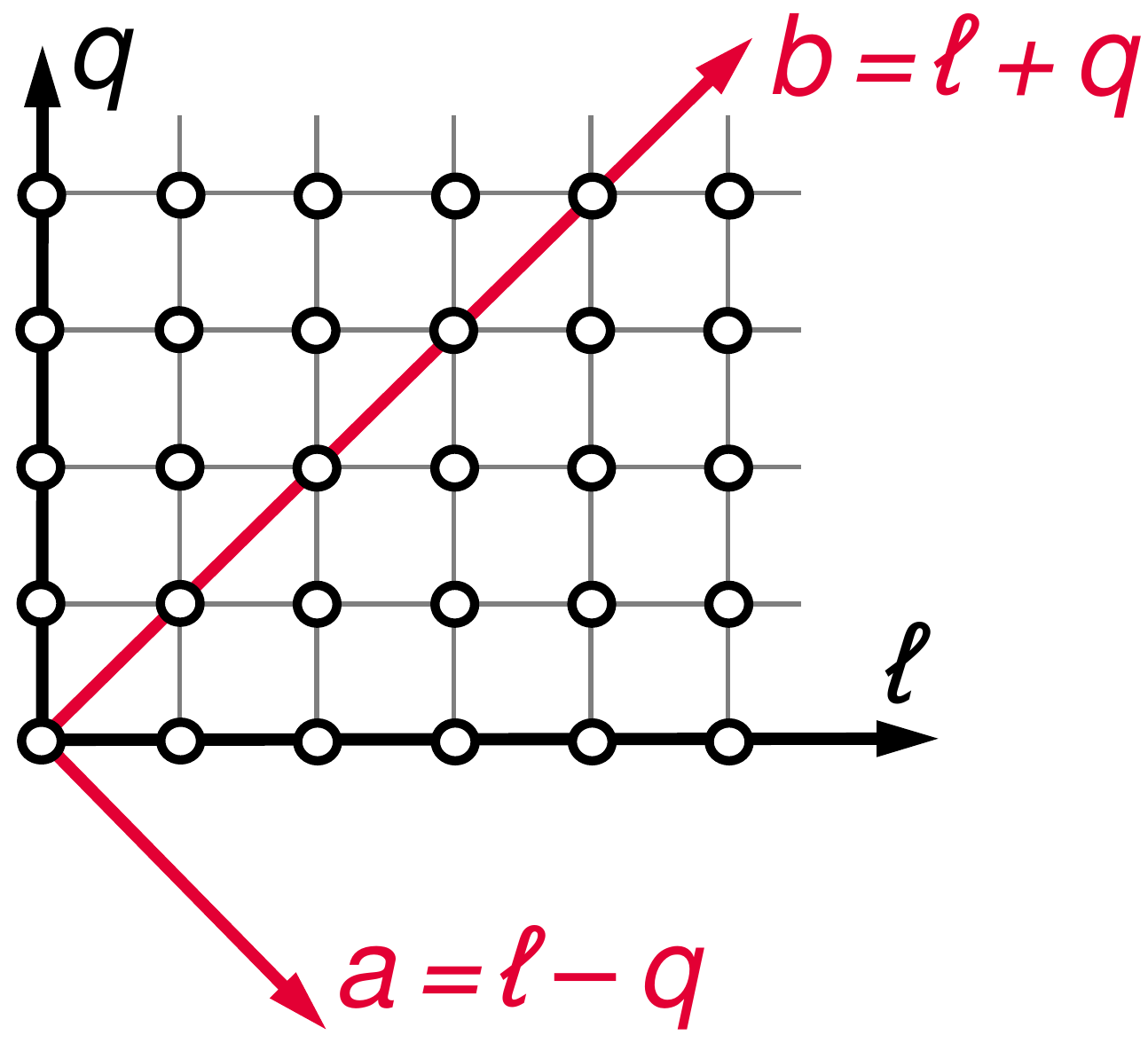}
    \caption{Transformation of indices for a double sum. }
   \label{fig:sum-index-transformation}
\end{figure}
\noindent The sum over index $k$ vanishes, except for $m-1-\ell-q=-p N$ with $p=0,1,2\ldots$, hence it equals $N\delta_{q,m-1-\ell+pN}$. We re-arrange the double sum over indices $\ell, q$ as a double sum over indices $a,b$ with $a=\ell-q, b=\ell+q$ (see Fig. \ref{fig:sum-index-transformation}). With $b=m-1+p N$ it follows 
\begin{equation}\label{eq:sum4}\begin{split}
& \sum_{p=0}^\infty -4 N \exp\left[\mathbbm{i}\left(\pi j\!-\!2x\right)\frac{m\!+\!pN}{N}\right] \sum_{a=-(m-1+pN),\Delta a=2}^{m-1+pN} \!\exp\left(\frac{-\mathbbm{i}\pi j a}{N}\right) ,
\end{split}\end{equation} 
whereby the index $a$ is incremented by $\Delta a=2$. We replace the index $a$ by $c$ via $a=2c-m+1-pN$. This gives 
\begin{equation}\label{eq:sum5}\begin{split}
& \sum_{a=-(m-1+pN),\Delta a=2}^{m-1+pN} \exp\left(\frac{-\mathbbm{i}\pi j a}{N}\right) \\ 
& = \exp\left(\mathbbm{i}\pi j\frac{m-1+pN}{N}\right) \sum_{c=0}^{m-1+pN} \exp\left(-2\pi\mathbbm{i}\frac{j}{N}c\right) \\ 
& = (-1)^{j p}\frac{\sin\left(\pi j m/N\right)} {\sin\left(\pi j/N\right)} .
\end{split}\end{equation} 
Inserting(\ref{eq:sum5}) into (\ref{eq:sum4}) and computing the infinite sum over index $p$ gives
\begin{equation}\label{eq:sum6}\begin{split}
2N\frac{\sin\left(\pi j m / N\right)}{\sin\left(\pi j / N\right)} \left(\mathbbm{i}\cot(x)-1\right)\exp\left(\mathbbm{i} m \frac{j\pi-2x}{N}\right) ,
\end{split}\end{equation}
which proves (\ref{eq:RMFoCD125}).

\end{document}